\documentclass[11pt]{article}
\usepackage{natbib,url,amsmath,amssymb,color,enumerate,adjustbox,array,fullpage,placeins,booktabs,xparse}

\begin{document}


\newcommand{\pa}{\mathrm{pa}}
\newcommand{\an}{\mathrm{an}}
\newcommand{\ch}{\mathrm{ch}}
\newcommand{\de}{\mathrm{de}}
\newcommand{\Cov}{\mathrm{Cov}}
\newcommand{\adj}{\mathrm{adj}}

\NewDocumentCommand{\rot}{O{45} O{1em} m}{\makebox[#2][l]{\rotatebox{#1}{#3}}}%

\newcommand{\nicolai}[1]{{\color{red}{Nicolai: {#1}}}}
\newcommand{\marloes}[1]{{\color{green}{Marloes: {#1}}}}
\newcommand{\christina}[1]{\textcolor{blue}{\textbf{Christina:
    }{\footnotesize #1}}}

\newcommand{\samp}{n}
\newcommand{\dims}{p}
\newcommand{\dfNoise}{df_{\varepsilon}}
\newcommand{\rhoNoise}{\rho_{\varepsilon}}
\newcommand{\strengthInt}{\sigma_{Z}}
\newcommand{\numberInt}{n_{I}}
\newcommand{\sparse}{p_{s}}
\newcommand{\snrPar}{\omega}
\newcommand{\strengthCycle}{w_{c}}
\newcommand{\nsims}{n_{sim}}
\newcommand{\nenv}{n_{\mathcal{E}}}
\newcommand{\env}{s}
\newcommand{\sampenv}{n_{\env}}
\newcommand{\Adj}{A}
\newcommand{\Adjtr}{B}
\newcommand{\subsamp}{k}
\newcommand{\Adjhist}{\mathbf{H}}
\newcommand{\query}{q}
\newcommand{\TPR}{{\mathrm{TPR}}}
\newcommand{\FPR}{{\mathrm{FPR}}}
\newcommand{\FNR}{{\mathrm{FNR}}}

\newcommand{\br}[1]{\left( {#1} \right)}
\newcommand{\nrm}[1]{\Vert {#1} \Vert}
\newcommand{\tr}{^{\top}}

\newcommand{\stararrow}{
  \setlength{\unitlength}{1mm}
  \begin{picture}(5,1)(0,0)
    \put(0.2,-1.1){*}
    \put(1,0){$\rightarrow$}
  \end{picture}
}

\newcommand{\arrowstar}{
  \setlength{\unitlength}{1mm}
  \begin{picture}(5,1)(0,0)
    \put(0.2,0){$\leftarrow$}
    \put(2.9,-.9){*}
  \end{picture}
}

\newcommand{\circarrow}{
  \setlength{\unitlength}{1mm}
  \begin{picture}(5,1)(0,0)
    \put(1,.85){\circle{1}}
    \put(1.2,0){$\rightarrow$}
  \end{picture}
}

\newcommand{\arrowcirc}{
  \setlength{\unitlength}{1mm}
  \begin{picture}(5,1)(0,0)
    \put(0.2,0){$\leftarrow$}
    \put(4.3,1){\circle{1}}
  \end{picture}
}

\newcommand{\circcirc}{
  \setlength{\unitlength}{1mm}
  \begin{picture}(6.55,1)(0,0)
    \put(1.3,1){\circle{1}}
    \put(1.75,1){\line(1,0){3.2}}
    \put(5.55,1){\circle{1}}
  \end{picture}
}

\newcommand{\circstar}{
  \setlength{\unitlength}{1mm}
  \begin{picture}(6.55,1)(0,0)
    \put(1.3,1){\circle{1}}
    \put(1.75,1){\line(1,0){3.2}}
    \put(4.6,-.75){*}
  \end{picture}
}

\newcommand{\tailstar}{
  \setlength{\unitlength}{1mm}
  \begin{picture}(5,1)(0,0)
    \put(0.4,1){\line(1,0){3.2}}
    \put(2.8,-.75){*}
  \end{picture}
}

\newcommand{\startail}{
  \setlength{\unitlength}{1mm}
  \begin{picture}(5,1)(0,0)
    \put(0.2,-1){*}
    \put(1.3,1){\line(1,0){3.2}}
  \end{picture}
}

\newcommand{\tailcirc}{
  \setlength{\unitlength}{1mm}
  \begin{picture}(5,1)(0,0)
    \put(0.4,.7){\line(1,0){3.2}}
    \put(4.2,.7){\circle{1}}
  \end{picture}
}

\newcommand{\circtail}{
  \setlength{\unitlength}{1mm}
  \begin{picture}(5,1)(0,0)
    \put(0, .7){\circle{1}}
    \put(0.4,.7){\line(1,0){3.2}}
  \end{picture}
}

\newcommand{\tailtail}{
  \setlength{\unitlength}{1mm}
  \begin{picture}(5,1)(0,0)
    \put(0.9,.7){\line(1,0){3.2}}
  \end{picture}
}

\title{Causal Structure Learning}

\author{Christina Heinze-Deml, Marloes H. Maathuis, and Nicolai
  Meinshausen \\
Seminar f\"ur Statistik, Department of Mathematics \\ ETH
  Zurich, Switzerland, CH-8092 Zurich \\ email: \{heinzedeml, maathuis, meinshausen\}@stat.math.ethz.ch}

\maketitle

\begin{abstract}
Graphical models can represent a multivariate distribution in a
convenient and accessible form as a graph. Causal models can be viewed
as a special class of graphical models that not only represent the
distribution of the observed system but also the distributions under
external interventions. They hence enable predictions under
hypothetical interventions, which is important for decision making. The challenging task of learning causal models from data always relies on some underlying assumptions. We discuss several recently proposed structure learning algorithms and their assumptions, and compare their empirical performance under various scenarios.
\end{abstract}

\tableofcontents

\section{INTRODUCTION}

A graphical model is a family of multivariate distributions associated with a graph, where the nodes in the graph represent random variables and the edges encode allowed conditional dependence relationships between the corresponding random variables \citep{Lauritzen1996}.
A \emph{causal} graphical model is a special type of graphical model, where edges are interpreted as direct causal effects. This interpretation facilitates predictions under arbitrary (unseen) interventions, and hence the estimation of causal effects \cite[e.g.,][]{wright34,Spirtes2000,Pearl2009}. This ability to make predictions under arbitrary interventions sets causal models apart from standard models.
We refer to \citet{didelez2017causalconcepts} for an introductory overview of causal concepts and graphical models.\footnote{Causal inference is also possible without graphs, using for example the Neyman-Rubin potential outcome model \citep[e.g.,][]{rubin2005causal}.
Single world intervention graphs (SWIGs)
\citep{richardson2013single} provide a unified framework for potential outcome and graphical approaches to causality.}

Structure learning is a model selection problem in which one estimates or learns a graph that
best describes the dependence structure in a given data set \citep{DrtonMaathuis17}. \emph{Causal} structure learning is the special case where one tries to learn the causal graph or certain aspects of it, and this is what we focus on in this paper. 
We describe various algorithms that have been developed for this purpose under different assumptions. We then compare the algorithms in a simulation study
to investigate their performances in settings where the assumptions
of a particular method are met, but also in settings where they are
violated.

The outline of the paper is as follows. Section \ref{sec: basic model}
discusses the basic causal model and its various assumptions. Section
\ref{sec: methods} describes different target graphical objects, such
as directed acyclic graphs or equivalence classes thereof, and
describes  algorithms that can learn them under certain assumptions.
Section \ref{sec:simulations} describes the simulation set-up, the
evaluation scheme, and the results. We close with a brief discussion in Section \ref{sec:discussion}.

\section{THE MODEL}\label{sec: basic model}

We formulate the model as a structural causal model \citep{Pearl2009}. In particular, we
consider a linear structural equation model \citep[e.g.,][]{Wright1921} for a $p$-dimensional random
variable $X=(X_1,\ldots,X_p)^t$ under noise contributions
$\varepsilon=(\varepsilon_1,\ldots, \varepsilon_p)^t$:
\begin{align}\label{eq:struceq}
  X_j & \leftarrow \sum_{k=1}^p  \beta_{j,k} X_k + \varepsilon_j  \qquad
  \mbox{ for  } j=1,\ldots,p,
\end{align}
or in vector notation,
\begin{align}\label{eq:struceq vector}
   X \leftarrow  B X + \varepsilon,
\end{align}
where $B$ is a $p \times p$ matrix with entries $B_{j,k}=\beta_{j,k}$. Thus, the distribution of $X$ is determined by the choice of $B$ and the distribution of $\varepsilon$.

This model is called \emph{structural} since it is interpreted as
the generating mechanism of $X$ (emphasized by the assignment operator
$\leftarrow$), where each structural equation is assumed to be
invariant to possible changes in the other structural equations. This
is also referred to as \emph{autonomy} \citep{Frisch38,Haavelmo1944}. This
assumption is key for causality, since it allows the derivation of the
distribution of $X$ under external interventions. For example, a gene knockout experiment
can be modeled by replacing the structural equation of the relevant gene, while keeping the other structural equations
unchanged. If the gene knockout experiment has significant off-target effects \citep[e.g.,][]{ChoEtAl14}, then this approach is problematic with respect to the autonomy assumption. A possible remedy consists of modeling the experiment as a simultaneous intervention on all genes that are directly affected by the experiment.

\subsection{Interventions}\label{subsec:interventions}

In this paper, we consider the following two types of interventions:
\begin{enumerate}[(a)]
\item A do-intervention (also called ``surgical" intervention): This intervention is modelled
  by replacing the structural equation
\[ X_j \leftarrow \sum_{k=1}^p  \beta_{j,k} X_k + \varepsilon_j
\quad\mbox{  by  } \quad X_j \leftarrow Z_j,\]
where $Z_j$ is the (either deterministic or random) value that
variable $X_j$ is forced to take in this intervention.
\item  An additive intervention (also called ``shift'' intervention): This intervention consists of adding additional noise,
  modelled by replacing the structural equation
\[ X_j \leftarrow \sum_{k=1}^p  \beta_{j,k} X_k + \varepsilon_j
\quad\mbox{ by  } \quad X_j \leftarrow \sum_{k=1}^p
\beta_{j,k} X_k + \varepsilon_j + Z_j,\]
where $Z_j$ is the additional noise (again either deterministic or
random) that is added to variable $X_j$. Shift interventions are
standard in the econometric literature on instrumental variables with
binary treatments where the additive shift of an exogenous instrument
changes the probability of a binary treatment variable
\citep{angrist1996identification}. Shift interventions are also
natural in biological settings where an inhibitor or enhancer can
amplify or decrease the presence of, for example, mRNA in a cell. If
the concentrations are amplified by a fixed factor, then this
corresponds to an additive shift in the log-concentrations.
\end{enumerate}

\subsection{Graphical representation} We can represent the model defined in \eqref{eq:struceq} as a directed graph $G$, where each variable $X_k$ is represented by a node $k$, $k=1,\ldots,p$, and there is an edge from node
$k$ to node $j$ ($k\neq j$) if and only if $\beta_{j,k} \neq 0$. Thus, the parents $pa(j,G)$ of node $j$ in $G$ correspond to the random variables that appear on the right hand side of the $j$th structural equation. In other words, $X_{\pa(j,G)}:=\{X_i: i\in pa(j,G)\}$ are the variables that are involved in the generating mechanism of $X_j$ and are also called the \emph{direct causes} of $X_j$ (with respect to $X_1,\dots,X_p$). In this sense, edges in $G$ represent direct causal effects and $G$ is also called a \emph{causal graph}. The nonzero $\beta_{j,k}$'s can be depicted as edge weights of $G$, yielding a weighted graph.
This weighted graph and the distribution of $\varepsilon$ fully determine the distribution of $X$.

The graph $G$ is called \emph{acyclic} if it does not contain a cycle\footnote{A cycle (sometimes also called directed cycle) is formed by a directed path from $i$ to $j$ together with the edge $j\to i$.}. A directed acyclic graph is also called a DAG. A directed graph is acyclic if and only if there is an ordering of the variables, called a \emph{causal order}, such that the matrix $B$ in equation \eqref{eq:struceq vector} is triangular. In terms of the
causal mechanism, acyclicity means that there are no feedback loops. We refer to
Section~\ref{sec: assumptions} for more details on cycles.

\subsection{Factorization and truncated factorization}\label{sec: fact and trunc}

If  $\varepsilon_1,\ldots,\varepsilon_p$
are jointly independent and $G$ is a DAG, then the probability density function $f(\cdot)$ of $X$ factorizes according to $G$:
\begin{align}\label{eq:factorization}
   f(x) = f(x_1,\dots,x_p)= \prod_{i=1}^p f(x_i|x_{pa(i,G)}).
\end{align}
Moreover, $f$ is then called \emph{Markov} with respect to $G$. This means that for pairwise disjoint subsets $A$, $B$ and $S$ of $V$ ($S = \emptyset$ is allowed) the following holds: if $A$ and $B$ are separated by $S$ in $G$ according to a graphical criterion called d-separation \citep{Pearl2009}, then $X_A$ and $X_B$ are conditionally independent given $X_S$ in $f$.

One can model
an intervention on $X_j$ by replacing the conditional density $f(x_j|x_{\pa(j)})$ by its conditional density under the intervention, keeping the other terms unchanged. For example, a do-intervention on $X_j$ yields the following factorization:
\begin{align*}
   f(x|do(x_j)) = g(x_j) \prod_{i=1, i\neq j}^p f(x_i |x_{\pa(i)}),
\end{align*}
where $g(\cdot)$ is the density of $Z_j$ (allowed to be a point mass). When intervening on several variables simultaneously, one simply conducts such replacements for all intervention variables. The resulting factorization is known as the g-formula \citep{Robins1986}, the manipulated density \citep{Spirtes2000}, or the truncated factorization formula \citep{Pearl2009}.

\subsection{Counterfactuals} We note that the structural causal
model is often discussed in the context of counterfactual outcomes. In
particular, if one assumes that $\varepsilon$ is identical under
different interventions, the model defines a joint distribution on all
possible counterfactual outcomes. The problematic aspect is clearly
that the realizations of the noise under different interventions can
never be observed simultaneously and any statement about the joint
distribution of the noise under different interventions is thus in
principle unfalsifiable and untestable
\citep{dawid2000causal}.
Without assuming anything on the joint noise distributions under
different interventions, a causal model can equivalently be
formulated via structural equations, a graphical model, or
potential outcomes
\citep{richardson2013single,imbens2014instrumental}.
For the causal structure learning methods discussed in this paper, no
assumption on the joint noise distribution is necessary and we chose
to use the structural equation framework for ease of exposition.

\subsection{Assumptions}\label{sec: assumptions}

We will consider various assumptions for the model defined by equation \eqref{eq:struceq vector}:\\

\noindent \textbf{Causal sufficiency.} Causal sufficiency refers to the absence of hidden (or latent) variables \citep{Spirtes2000}. There are two common options for the modeling of
hidden variables\footnote{In this manuscript we look at the behavior of various methods under the presence and
absence of latent confounding. Throughout, we do not allow hidden selection
variables, that is, unmeasured variables that determine if a unit is included in the data sample. More details on selection variables can be found in, e.g., \cite{SpirtesMeekRichardson99}.\label{footnote: selection vars}}: They can be modeled
explicitly as nodes in the
the structural equations, or they can manifest themselves as a
dependence between the noise terms
$(\varepsilon_1,\ldots,\varepsilon_p)$, where the noise terms are
assumed to be independent in the absence of latent confounding. 

\smallskip
\noindent \textbf{Causal faithfulness.} 
We saw in Section \ref{sec: fact and trunc} that the distribution of $X$ generated from equation \eqref{eq:struceq vector} is Markov with respect
to the causal DAG, meaning that if $A$ and $B$ are d-separated
by $S$ in the causal DAG, then $X_A$ and
$X_B$ are conditionally independent given $X_S$. The reverse implication is called causal faithfulness. Together, the causal Markov and causal faithfulness assumptions imply that d-separation relationships in the causal DAG have a one-to-one correspondence with conditional independencies in the distribution.

\smallskip
\noindent \textbf{Acyclicity.}
Cycles can be used to model instantaneous feedback mechanisms. In the
presence of cycles, the structural equations~\eqref{eq:struceq} are
typically interpreted (implicitly) as a dynamical system.
There are various assumptions that can be made about the strength of
cycles in the graph\footnote{We exclude self-loops (an edge from a node to itself),
  as models would
  be unidentifiable if self-loops were allowed  \citep[see, e.g.,][]{rothenhausler2015backshift}.}:
\begin{enumerate}[(i)]
\item Existence of a unique equilibrium solution of equation \eqref{eq:struceq
    vector}. Is there a unique solution $X$ for each realization
  $\varepsilon$ such that $X=BX+\varepsilon$ or, equivalently,
  $(I-B)X=\varepsilon$, where $I$ is the $p$-dimensional identity matrix?  Existence of a unique
  equilibrium requires that  $I-B$  is
invertible. In this case the equilibrium is \[ X  = (I-B)^{-1}\varepsilon.\]
\item Convergence to a stable equilibrium. Iterating equation~\eqref{eq:struceq
    vector} from any starting
  value  $X^{(0)}$ for $X$ (and for a fixed and constant realization of the noise
  $\varepsilon$), we can form an iteration $X^{(k)}= B
  X^{(k-1)}+\varepsilon$ for $k\in \mathbb{N}$. The question is then
  whether the iterations converge to the equilibrium, that is, whether
  $\lim_{k\to\infty} X^{(k)} =  (I-B)^{-1}\varepsilon$. This
  convergence requires that the spectral radius of $B$ is
smaller than 1.
\item Existence of a stable equilibrium under do-interventions. This requires in addition that the
cycle product (the maximal product of the
coefficients along all loops in the graph)  is smaller  than 1, see for example
\cite{rothenhausler2015backshift}.
\end{enumerate}
DAGs fulfil all three assumptions (i)-(iii) above trivially as their spectral radius
and cycle product both vanish identically.

\smallskip
\noindent \textbf{Gaussianity of the noise distribution.} We
consider both Gaussian distributions and t-distributions with various degrees of freedom.

\smallskip
\noindent \textbf{One or several experimental settings.} 
We consider both homogeneous data, where all observations are from the same experimental setting, and heterogeneous data, where the observations come from
different experimental settings. In particular, we consider settings with unknown shift-interventions and known do-interventions. 

\smallskip
\noindent \textbf{Linearity.} 
While the assumptions and the models have been discussed in the
context of linear models, the ideas can be extended to nonlinear
models and to discrete random variables to various degrees.

\section{METHODS}\label{sec: methods}

Since different structure learning methods output different types of graphical objects, we first discuss the various target graphical objects in Section \ref{sec: target graphical objects}. To conduct a comparison based on such different graphical targets, we focus on certain ancestral relationships that can be read off from all objects (see Section \ref{sec: ancestral and parental relationships}). The different algorithms and their assumptions are discussed in Section \ref{sec: considered methods}, and their assumptions are summarized in Table \ref{tab:assume}.

\subsection{Target graphical objects}\label{sec: target graphical objects}

The structure learning methods that we will compare use different types of data, from purely
observational data to data with clearly labelled interventions, from not allowing hidden variables and cycles to allowing both of these. As a result,
the different methods learn the underlying causal graph at different
levels of granularity. At the finest level of granularity, a
method learns the underlying \emph{directed graph} (DG)
from equation~\eqref{eq:struceq}.
If the method assumes acyclicity (no feedback), then the
target object is a \emph{directed acyclic graph} (DAG).

Under the model of equation~\eqref{eq:struceq vector} with acyclicity, independent and multivariate Gaussian errors and i.i.d.\ observational data, the underlying causal DAG is generally not identifiable. Instead, one can identify the Markov equivalence class of DAGs, that is, the set of DAGs that encode the same set of d-separation relationships \citep{Pearl2009}.
A Markov equivalence can be conveniently summarized by another
graphical object, called a  \emph{completed  partially directed
  acyclic graph} (CPDAG) \citep{Andersson1997, Chickering02-CPDAG}. A CPDAG can be interpreted as follows: $i \to j$ is in the CPDAG if $i \to j$ in every DAG in the Markov equivalence class, and $i \circcirc j$ in the CPDAG if there is a DAG with $i \to j$ and a DAG with $i \leftarrow j$ in the Markov equivalence class. Thus, edges of the type $\circcirc$ represent uncertainty in the edge orientation.

DAGs are not closed under marginalization. In the presence of latent variables,
some algorithms
therefore aim to learn a
different object, called a \emph{maximal ancestral graph} (MAG)
\citep{Richardson2002}. In general, MAGs contain three types
of edges: $i \tailtail j$, $i \to j$ and $i \leftrightarrow j$, but in our
settings without selection variables (see footnote \ref{footnote: selection vars}), $i \tailtail j$ does not occur.
A MAG encodes conditional independencies via
m-separation \citep{richardson2002ancestral}. Every DAG with latent
variables can be uniquely mapped to a MAG that encodes the same
conditional independencies and the same ancestral relationships among
the observed variables. Ancestral relationships can be read off from
the edge marks of the edges: a tail mark $i \tailstar j$ means that $i$ is an
ancestor of $j$ in the underlying DAG, and an arrowhead $i \arrowstar j$ means that $i$ is not an ancestor of $j$ in the underlying DAG, where $*$ represents any of the possible edge marks (again assuming no selection variables).

Several MAGs can encode the same set of conditional independence
relationships. Such MAGs form a Markov equivalence class, which can be
represented by a \emph{partial ancestral graph} (PAG) \citep{Richardson2002,AliRichardsonSpirtes09}. A PAG can
contain the following edges: $i \to j$, $i \tailtail j$, $i \tailcirc j$, $i \leftrightarrow j$, $i \circarrow j$, and $i \circcirc j$,
but the edges $i \tailtail j$ and $i \tailcirc j$ do not occur in our setting
without selection variables. The interpretation of the edge marks is as
follows. A tail mark means that this tail mark
is present in all MAGs in the Markov equivalence class, and an
arrowhead means that
this arrowhead is present in all MAGs in the Markov equivalence class. A circle mark
represents uncertainty about the edge mark, in the sense that there is a MAG in the Markov equivalence class where this
edge mark is a tail, as well as a MAG where this edge mark is an
arrowhead.

\subsection{Ancestral and parental relationships}\label{sec: ancestral and parental relationships}

To compare methods that output the different graphical objects discussed above, we
focus on the following two basic questions for any variable $X_j$, $j\in \{1,\dots,p\}$, and the
underlying causal DAG $G$:
\begin{enumerate}[(a)]
\item What are the direct causes of $X_j$, or equivalently, what is $\pa_G(j)$? The parents are
  important, since they completely determine the distribution of $X_j$. Hence, the conditional
  distribution $X_j|X_{\pa(j)}$ is constant, even under arbitrary
  interventions on subsets of $X_{\{1,\dots,p\}\setminus\{j\}}$.
  The set of parents is unique in this respect and allows to
  make accurate predictions about $X_j$ even under arbitrary interventions on all
  other variables.
  Moreover, the (possible) parents of $X_j$ can be used to estimate (bounds on) the total causal effect of $X_j$ on any of the other variables \citep{Maathuis2009,Maathuis2010,Stekhoven2012,Nandy2014}.
\item What are the causes of $X_j$, or equivalently, what is the set of ancestors $\an_G(j)$ (the set of nodes from which there is a directed path to $j$ in $G$)? The ancestors are important, since any intervention on ancestors of $X_j$ has an effect on the
  distribution of $X_j$, as long as no other do-type interventions
  happen along the path. Thus, if we want to manipulate the distribution of $X_j$, we can consider interventions on subsets of $X_{\an_G(j)}$.
\end{enumerate}

\subsection{Considered methods}\label{sec: considered methods}

We include at least one algorithm from each of the following five main classes of causal structure learning algorithms: constraint-based methods, score-based methods, hybrid methods, methods based on structural equation models with additional restrictions,
and methods exploiting invariance properties.
Limiting ourselves to algorithms with an implementation in \textsf{R} \citep{R}, we obtain the following selection of methods, with assumptions summarized in Table \ref{tab:assume}:
\begin{itemize}
   \item Constraint-based methods: PC \citep{Spirtes2000}, rankPC \citep{harrisdrton13}, FCI \citep{Spirtes2000}, and rankFCI\footnote{rankFCI is obtained by using rank correlations in FCI, analogously to rankPC.}
   \item Score-based methods: GES \citep{Chickering2002}, rankGES \citep{NandyEtAl17}, GIES \citep{Hauser2012}, and rankGIES\footnote{rankGIES is obtained by using rank correlations in GIES, analogously to rankGES.}
   \item Hybrid methods: MMHC \citep{Tsamardinos2006}
   \item Structural equation models with additional restrictions: LINGAM \citep{Shimizu2006}
   \item Exploiting invariance properties: BACKSHIFT \citep{rothenhausler2015backshift}
\end{itemize}

We have not included methods for time series data,
mixed data, or Bayesian methods. Other excluded methods that make use of interventional data include \citet{cooper1999causal, Tian2001} and \citet{Eaton2007}, where the latter does not require knowledge of the precise location of interventions in a similar spirit to \cite{rothenhausler2015backshift}. \citet{Hyttinen2012} also makes use of intervention data to learn feedback models, assuming do-interventions, while \cite{peters2015causal} permits to build a graph nodewise by estimating the parental set of each node separately.

\begin{table}
\tabcolsep7.5pt
\caption{The assumptions (see Section \ref{sec: assumptions}) and output format for the different
  methods. (For example, PC requires acyclicity, causal faithfulness and causal sufficiency, and LINGAM requires non-Gaussian errors.)
  Please note that linearity is not explicitly listed, but all versions of the algorithms based on rank-correlations allow certain types of nonlinearities.
  The different output formats are: DG (directed
  graph), DAG (directed acyclic graph), PDAG (partially directed
  acyclic graph), CPDAG (completed partially directed graph) and PAG
  (partial ancestral graph). }
\label{tab:assume}
\begin{center}

\begin{tabular}{rccccccc} &\rot{(rank)PC}&\rot{(rank)FCI}&\rot{(rank)GES}&\rot{(rank)GIES}&\rot{MMHC}&\rot{LINGAM}&\rot{BACKSHIFT} \\ \hline
Causal sufficiency  & yes & no & yes & yes & yes  & yes  & no \\
Causal faithfulness & yes & yes & yes & yes & yes & no & no \\
Acyclicity & yes & yes& yes & yes & yes & yes & no \\
Non-Gaussian errors & no & no  & no & no  & no & yes &  no\\
Unknown shift interventions  & no & no & no & no  & no & no
& yes  \\
Known do-interventions & no & no & no & yes & no & no & no
\\
\hline
&&&&&&& \\
Output & \rot{CPDAG} &\rot{PAG}&\rot{CPDAG}&\rot{PDAG}&\rot{DAG}&\rot{DAG}&\rot{DG}
\end{tabular}

\end{center}
\end{table}

\subsubsection{(rank)PC and (rank)FCI}

The PC algorithm \citep{Spirtes2000} is named after its inventors Peter Spirtes and Clark Glymour. It is a constraint-based algorithm that assumes acyclicity, causal faithfulness and causal sufficiency. It conducts numerous conditional independence tests to learn about the structure of the underlying DAG. In particular, it learns the CPDAG of the underlying DAG in three steps: (i) determining the skeleton, (ii) determining the v-structures, and (iii) determining further edge orientations. The skeleton of the CPDAG is the undirected graph obtained by replacing all directed edges by undirected edges. The PC algorithm learns the skeleton by starting with a complete undirected graph. For $k=0,1,2,\dots$ and adjacent nodes $i$ and $j$ in the current skeleton, it then tests conditional independence of $X_i$ and $X_j$ given $X_S$ for all   $S \subseteq \adj(i)\setminus \{j\}$ with $|S|=k$, and for all $S\subseteq \adj(j)\setminus\{i\}$ with $|S|=k$.  The algorithm removes an edge if a conditional independence is found (that is, the null hypothesis of independence was not rejected at some level $\alpha$), and stores the corresponding separating set $S$. Step (i) stops if the size of the conditioning set $k$ equals the degree of the graph.

In step (ii), all edges are replaced by $\circcirc$, and the algorithm considers all unshielded triples, that is, all triples $i \circcirc j \circcirc k$ where $i$ and $k$ are not adjacent. Based on the separating set that led to the removal of $i \tailtail k$, the algorithm determines whether the triple should be oriented as a v-structure $i \to j \leftarrow k$ or not. Finally, in step (iii) some additional orientation rules are applied to orient as many of the remaining undirected edges as possible.

The PC algorithm was shown to be consistent in certain sparse high-dimensional settings \citep{Kalisch2007}. There are various modifications of the algorithm. We use the order-independent version of \cite{ColomboMaathuis14}. The PC algorithm does not impose any distributional assumptions, but conditional independence tests are easiest in the binary and multivariate Gaussian settings. \cite{harrisdrton13} proposed a version of the PC algorithm for certain Gaussian copula distributions. We include this algorithm in our comparison and denote it by rankPC. There is also a version of the PC algorithm that allows cycles \citep{Richardson1999}, but we did not find an \textsf{R} implementation of it.

The Fast Causal Inference (FCI) algorithm \cite{SpirtesMeekRichardson99,Spirtes2000} is a modification of the PC algorithm that drops the assumption of causal sufficiency by allowing arbitrarily many hidden variables. The output of the FCI algorithm can be interpreted as a PAG \citep{Zhang08-causal-reasoning-ancestral-graphs}. The first step of the FCI algorithm is the same as step (i) of the PC algorithm, but the FCI algorithm needs to conduct additional tests to learn the correct skeleton. There are also additional orientation rules, which were shown to be complete in \cite{Zhang08-orientation-rules}. Since the additional tests can slow down the algorithm considerably, faster adaptations have been developed, such as RFCI \citep{Colombo2012} and FCI+ \citep{Claassen2013}. \cite{Colombo2012} proved high-dimensional consistency of FCI and RFCI. The idea of \cite{harrisdrton13} can also be applied to FCI, leading to rankFCI.

\subsubsection{(rank)GES and (rank)GIES}

Greedy equivalence search (GES) \citep{Chickering2002} is a score-based algorithm that assumes acyclicity, causal faithfulness and causal sufficiency. Score-based algorithms use the fact that each DAG can be scored in relation to the data, typically using a penalized likelihood score. The algorithms then search for the DAG or CPDAG that yields the optimal score. Since the space of possible graphs is typically too large, greedy approaches are used. In particular, GES learns the CPDAG of the underlying causal DAG by conducting a greedy search on the space of possible CPDAGs. Its greedy search consists of a forward phase, where it conducts single edge additions that yield the maximum improvement in score, and then a backward phase, where it conducts single edge deletions. Despite the greedy search, \cite{Chickering2002} showed that the algorithm is consistent under some assumptions (for fixed $p$). \cite{NandyEtAl17} showed high-dimensional consistency of GES.

Greedy interventional equivalence search (GIES) \citep{Hauser2012} is an adaptation of GES to settings with data from different known do-interventions. Due to the additional information from the interventions, its target graphical object is a so-called interventional Markov equivalence
class, which is a sub-class of the Markov equivalence class of the
underlying DAG and can be seen as a partially directed acyclic graph (PDAG).

\cite{NandyEtAl17} showed a close connection between score-based and constraint-based methods for multivariate Gaussian data. As a result, the copula methods that can be used for the PC and FCI algorithms, can be transferred to the GES and GIES algorithms. We include these algorithms in our comparison, and refer to them as rankGES and rankGIES.

\subsubsection{MMHC}

Max-Min Hill Climbing (MMHC) \citep{Tsamardinos2006} is a hybrid algorithm that assumes acyclicity, causal faithfulness, and causal sufficiency. Hybrid algorithms combine ideas from both constraint-based and score-based approaches. In particular, MMHC first learns the CPDAG skeleton using the constraint-based Max-Min Parents and Children (MMPC) algorithm and then performs a score-based hill-climbing DAG search to determine the edge orientations. Its output is a DAG. \cite{NandyEtAl17} showed that the algorithm is not consistent for fixed $p$, due to the restricted score-based phase.

\subsubsection{LINGAM}

LINGAM \citep{Shimizu2006} is an acronym derived from ``linear non-gaussian acyclic models''
and has been designed for model
\eqref{eq:struceq vector} with non-Gaussian noise. It assumes acyclicity and
causal sufficiency. It is based on the fact that $X = A \varepsilon$
with $A = (I-B)^{-1}$. This can be viewed as a source separation
problem, where identification of the matrix $B$ is equivalent to
identification of the mixture matrix $A$. It was shown in
\cite{Comon94} that whenever at most one of the components
of $\varepsilon$ is Gaussian, the mixing matrix is identifiable up to scaling and permutation
of columns, via independent component analysis (ICA). This observation lies at the basis of the LINGAM method.
There are various modifications of LINGAM, for example to allow for
hidden variables \citep{Hoyer2008b} or cycles
\citep{lacerda2012discovering}. There is also a different
implementation called DirectLINGAM \citep{Shimizu2011} that uses a pairwise causality measure instead of ICA. Since only ICA-based LINGAM assuming acyclicity and causal sufficiency is available in \textsf{R}, we include this version in our comparison.

\subsubsection{BACKSHIFT}

BACKSHIFT \citep{rothenhausler2015backshift} makes use of
non-i.i.d.\ structure in the data and unknown shift interventions on
variables.
Assume that the data are divided into distinct blocks
$\mathcal{E}$. Let $\Gamma_e\in \mathbb{R}^{p \times p}$ be the
empirical Gram
matrix of the $p$ variables in block $e\in \mathcal{E}$ of the
data. In the absence of shift-interventions the expected values of
$\Gamma_e$ would be identical for all $e\in \mathcal{E}$. Under
unknown-shift interventions the Gram matrices can change from block to
block. However, for the true matrix $B$ of causal coefficients
from equation~\eqref{eq:struceq vector},
it can be shown that the expected value of
\[ (I-B) (\Gamma_e - \Gamma_{e'} ) (I-B)^t \]
is a diagonal matrix for all $e,e'\in \mathcal{E}$, even in the
presence of latent confounding.
BACKSHIFT estimates $I-B$ (and hence $B$) by a joint diagonalization
of all Gram differences $\Gamma_e - \Gamma_{e'}$ for all pairs $e,e'\in
\mathcal{E}$. A necessary and sufficient condition for
identifiability of the causal matrix $B$ is as
follows. Let $\eta_{e,k}$ be the variance of the noise interventions
at variable $k\in\{1,\ldots,p\}$ in setting $e\in\mathcal{E}$.
Full identifiability requires that we can find for each pair of
variables $(k,l)$ two settings $e,e'\in\mathcal{E}$ such that the
product $\eta_{e,k}\eta_{e',l}$ is \emph{not} equal to the product
$\eta_{e,l},\eta_{e',k}$. A consequence of this necessary and
sufficient condition for identifiability is $|\mathcal{E}|\ge 3$, that
is, we need to observe at
least three different blocks of data for identifiability.

\section{EMPIRICAL EVALUATION}\label{sec:simulations}

We conducted an extensive simulation study  to evaluate and compare the methods,  paying
particular attention to sensitivity of the methods to model violations.
We are also interested in realistic boundaries (in terms of the number of variables, the sample size, and other simulation parameters) beyond which we cannot expect a reasonable reconstruction of the underlying graph.

In Section \ref{subsec:dat_gen}, we describe the data generating mechanism used in the simulation study.  Section \ref{subsec:eval} discusses the framework for comparison of the considered methods, and Section \ref{subsec:evalresults} contains the results.

The code is available in the \textsf{R} package \texttt{CompareCausalNetworks} \citep{CCN-r-pkg} along with further documentation. All methods are called through the interface offered by the \texttt{CompareCausalNetworks} package which depends on the packages \texttt{backShift} \citep{backshift-pkg}, \texttt{bnlearn} \citep{bnlearn-pkg} and \texttt{pcalg} \citep{KalischEtAl12}  for the code of the considered methods. In particular, BACKSHIFT is in \texttt{backShift}, MMHC is in \texttt{bnlearn}, and all other considered methods are in \texttt{pcalg}.

\subsection{Data generation}\label{subsec:dat_gen}
We generate data sets that differ with respect to the following characteristics: the number of observations $\samp$, the number of variables $\dims$, the expected number of edges in $\Adjtr$, the noise distribution, the correlation of the noise terms, the type, strength and number of interventions, the signal-to-noise ratio, the presence and strength of a cycle in the graph, and possible model misspecifications in terms of nonlinearities. The function \texttt{simulateInterventions()} from the package \texttt{CompareCausalNetworks} implements the simulation scheme that we describe in more detail below.

We first generate the adjacency matrix $\Adjtr$. Assume the variables with indices $\{1, \ldots, p \}$ are causally ordered. For each pair of nodes $i$ and $j$, where $i$ precedes $j$ in the causal ordering, we draw a sample from $\text{Bern}(\sparse)$ to determine  the presence of an edge from $i$ to $j$. After having sampled the non-zero entries of $\Adjtr$ in this fashion, we sample their corresponding coefficients from $\text{Unif}(-1,1)$. As described below, the edge weights are later rescaled 
to achieve a specified signal-to-noise ratio. We exclude the possibility of $\Adjtr = \mathbf{0}$, that is, we resample until $\Adjtr$ contains at least one non-zero entry.

Second, we simulate the interventions. We let $\numberInt$ denote the
total number of (interventional and observational) settings that are
generated. Let $I\in\{0,1\}^{\numberInt\times p}$ be an indicator
matrix, where an entry $I_{e,k}=1$ indicates that variable $k$ is
intervened on in setting $e$ and a zero entry indicates that no
intervention takes place. For each variable $k$, we first set the
$k$-th column
$I_{\cdot k}\equiv 0$ and then sample one setting $e'$ uniformly at
random and set  $I_{e'k}=1$.  In other words, each variable is
intervened on in exactly one setting. It is possible that there are
settings where no interventions take place, corresponding to
 zero rows of the matrix $I$, representing the observational setting.
 Similarly, there may be settings where interventions are performed on multiple variables at once. After defining the settings, we sample (uniformly at random with replacement) what setting each data point belongs to. So for each setting we generate approximately the same number of samples. In any generated data set, the interventions are all of the same type, that is, they are either all shift or all do-interventions (with equal probability). In both cases, an intervention on a variable $X_j$ is modeled by sampling $Z_j$ from a t-distribution as $Z_j \sim\strengthInt \cdot t(\dfNoise)$ (cf.\ Section~\ref{subsec:interventions}). If $\strengthInt = 0$ is sampled, it is taken to encode that no interventions should be performed. In that case, all interventional settings correspond to purely observational data.

Third, the noise terms $\varepsilon$ are generated by first sampling from a $p$-dimensional zero-mean Gaussian distribution with covariance matrix $\Sigma$,
where ${\Sigma}_{i,i} = 1$ and ${\Sigma}_{i,j} = \rhoNoise$. The magnitude of $\rhoNoise$ models the presence and the strength of hidden variables (cf.\ Section~\ref{sec: assumptions}). For a positive value of $\rho$ the correlation structure corresponds to the presence of a hidden variable that affects each observed variable.
To steer the signal-to-noise ratio, we set the variance of the noise
terms of all nodes except for the source nodes to $\snrPar$, where $0 <
\snrPar \leq 1$. Stepping through the variables in causal order, for
each variable $X_j$ that has parents, we uniformly rescale the edge
weights $\beta_{j,k}$ in the $j$th structural
equation such that the variance of the sum $\sum_{k=1}^p  \beta_{j,k}
X_k + \varepsilon_j$ is approximately equal to one in the observational
setting. In other words, the parameter $\snrPar$ steers what
proportion of the variance stems
from the
noise $ \varepsilon_j$. The signal-to-noise ratio can then be computed
as $\text{SNR} = (1-\snrPar)/\snrPar$ (in the absence of hidden
variables).

Fourth, if the causal graph shall contain a cycle, we sample two nodes $i$ and $j$ such that adding an edge between them creates a cycle in the causal graph. We then compute the coefficient for this edge such that the cycle product is 1. Subsequently, we sample the sign of the coefficient with equal probability and set the magnitude by scaling the coefficient by $\strengthCycle$, where $0 < \strengthCycle < 1$.

Fifth, we transform the noise variables to obtain a t-distribution with $\dfNoise$ degrees of freedom. $X$ is then generated as $X  = (I-\Adjtr)^{-1}\varepsilon$ in the observational case; under a shift interventions $X$ can be generated as $X  = (I-\Adjtr)^{-1}(\varepsilon + Z)$ where the coordinates of $Z$ are only non-zero for the variables that are intervened on. Under a do-intervention on $X_j$, $\beta_{j,k}$ for $k = 1, \ldots, p$ are set to 0 to yield $\Adjtr'$ and $\varepsilon_j$ is set to $Z_j$ to yield $\varepsilon_j'$. We then obtain $X$ as $X  = (I-\Adjtr')^{-1}\varepsilon'$.

Sixth, if nonlinearity is to be introduced, we marginally transform all variables as $X_j \leftarrow \text{tanh}(X_j)$.

Lastly, we randomly permute the order of the variables in $X$ before running the algorithms. Methods that are order-dependent can therefore not exploit any potential advantage stemming from a data matrix with columns ordered according to the causal ordering or a similar one.

\subsubsection{Considered settings}
We sample the simulation parameters uniformly at random from the following sets.
\begin{itemize}
	\item[--] Sample size $\samp \in \{500,2000,5000,10000 \}$
	\item[--] Number of variables $\dims \in \{2, 3,4,5,6,7,8,9, 10, 12, 15, 20,  50, 100 \}$
	\item[--] Edge density parameter $\sparse \in \{0.1,0.2,0.3,0.4 \}$
	\item[--] Number of interventions $\numberInt \in \{3,4,5 \}$
	\item[--] Strength of the interventions $\strengthInt \in \{0, 0.1,0.5,1,2,3,5,10 \}$
	\item[--] Degrees of freedom of the noise distribution $\dfNoise \in \{2,3,5,10,20,100 \}$
	\item[--] Strength of hidden variables $\rhoNoise \in \{0,0.1,0.2,0.5,0.8 \}$
	\item[--] Proportion of variance from noise $\snrPar \in \{0.1, 0.2, 0.3, 0.4, 0.5, 0.6, 0.7, 0.8, 0.9 \}$
	\item[--] Strength of cycle $\strengthCycle \in \{0.1,0.25,0.5,0.75,0.9 \}$
\end{itemize}
In total, we consider 842 different configurations. For each sampled configuration, we generate 20 different causal graphs with one data set per graph. Appendix~\ref{app:subsubsec:sim_settings} summarizes the number of simulation settings for different values of the simulation parameters.

\subsection{Evaluation methodology}\label{subsec:eval}
As the targets of inference differ between the considered methods, we evaluate a method's accuracy for recovering (a) parental and (b) ancestral relations (see also Section \ref{sec: ancestral and parental relationships}). For each of these, we look at a method's performance for predicting (i) the existence of a relation, (ii) the absence of a relation and (iii) the potential existence of a relation. We formulate these different categories as so-called queries which are further described in Section \ref{subsubsec:queries}.

An additional challenge in comparing a diverse set of methods
involves choosing the options and the proper amount of regularization
that determines the sparsity of the estimated structure. We address
this challenge in two ways. First, we run different configurations of
each method's tuning parameters and options as detailed in the
Appendix in Section \ref{app:subsubsec:methods_pars}. In the
evaluation of the methods for a certain metric, we choose the method's
configuration that yielded the best results under the considered
metric in each simulation setting (averaged over the twenty data sets
for each setting). This means that the results are optimistically
biased, but we found that the ranking was largely insensitive to the tuning parameter
choices.
Secondly, we use a subsampling scheme (stability ranking) so that each method outputs a ranking of pairs of nodes for a given query. For instance, the first entry in this ranking for the existence of parental relations is the edge most likely to be present in the underlying DAG.
Further details are given in Sections \ref{subsubsection:ranking} and \ref{subsubsection:metrics}.

\subsubsection{Considered queries}\label{subsubsec:queries}
For both the parental and the ancestral relations, we consider three queries. The existence of a relation is assessed by the queries \texttt{isParent} and \texttt{isAncestor}; the absence of a relation is assessed by the queries \texttt{isNoParent} and \texttt{isNoAncestor}; the potential existence of a relation is assessed by the queries \texttt{isPossibleParent} and \texttt{isPossibleAncestor}.

All queries return a connectivity matrix which we denote by $\Adj$. The interpretation of the entries of $\Adj$ differ according to the considered query:

\paragraph*{Parental relations}

\begin{enumerate}
	\item \texttt{isParent}
	This query cannot be easily answered by methods that return a PAG.
For the other graphical objects, $\Adj_{i,j} = 1$ if $i\to j$ in the estimated graph, and $\Adj_{i,j}=0$ otherwise.
		
	\item \texttt{isPossibleParent}
	Entry $\Adj_{i,j} = 1$ if there is an edge of type $i \tailstar j$ or $i \circstar j$
	in the estimated graph.
	Concretely, for methods estimating DGs or DAGs $\Adj_{i,j} = 1$ if $i\to j$ in the estimated graph, for PDAGs and CPDAGs $\Adj_{i,j} = 1$ if $i\to j$ or $i\circcirc j$ in the estimated graph, and for PAGs $\Adj_{i,j} = 1$ if $i\to j$, $i \tailcirc j$,  $i \tailtail j$, $i \circarrow j$, $i \circcirc j$ or $i \circtail j$ in the estimated graph.
	Otherwise, $\Adj_{i,j} = 0$.
		
	\item \texttt{isNoParent} The complement of the query \texttt{isPossibleParent}: If the latter returns the connectivity matrix $\Adj'$, then entry $\Adj_{i,j} = 1$ if  $\Adj'_{i,j} = 0$ and $\Adj_{i,j} = 0$ if  $\Adj'_{i,j} = 1$.
\end{enumerate}

\paragraph*{Ancestral relations}
\begin{enumerate}
			
	\item \texttt{isAncestor} Entry $\Adj_{i,j} = 1$ if there is a path from $i$ to $j$ with edges of type $\tailstar$. For example, for DGs, DAGs and CPDAGs this reduces to a directed path. Otherwise, $\Adj_{i,j} = 0$.
		
	\item \texttt{isPossibleAncestor} Entry $\Adj_{i,j} = 1$ if there is a path from $i$ to $j$ such that no edge on the path points towards $i$ (possibly directed path), and $\Adj_{i,j} = 0$ otherwise.
In general, such a path can contain edges of the type $i \tailstar j$ and $i \circstar j$. For DAGs and DGs this again reduces to a directed path, and for CPDAGs it is path with edges $\circcirc$ and $\to$.

	\item \texttt{isNoAncestor} The complement of the query \texttt{isPossibleAncestor}: If the latter returns the connectivity matrix $\Adj'$, then entry $\Adj_{i,j} = 1$ if  $\Adj'_{i,j} = 0$ and $\Adj_{i,j} = 0$ if ~$\Adj'_{i,j} = 1$.
\end{enumerate}

\subsubsection{Stability ranking}\label{subsubsection:ranking}
To obtain a ranking of pairs of nodes for a given query, we run the method under consideration on $\nsims = 100$ random subsamples of the data, where each subsample contains approximately $n/2$ data points. More specifically, we use the following stratified sampling scheme: In each round, we draw samples from $1/\sqrt{2}\cdot\numberInt$ settings, where $\numberInt$ denotes the total number of (interventional and observational) settings. In each chosen setting $\env$, we sample $1/\sqrt{2}\cdot\sampenv$ observations uniformly at random without replacement, where $\sampenv$ denotes the number of observations in setting $\env$. After a random permutation of the order of the variables, we run the method on each subsample
and evaluate the method's output with respect to the considered query.

For each subsample $\subsamp$ and a particular query $\query$, we obtain the corresponding connectivity matrix $\Adj$. We can then rank all pairs of nodes $i,j$ according to the frequency $\pi_{i,j}\in[0,1]$ of the occurrence of $\Adj_{i,j} = 1$ across subsamples. Ties between pairs of variables can be broken with the results of the other queries---for instance, if the query is \texttt{isParent}, ties are broken with counts for \texttt{isPossibleParent}. This {stability ranking} scheme is implemented in the function \texttt{getRanking()} in the package \texttt{CompareCausalNetworks}. Further details about the tie breaking scheme are given in the package documentation.

\subsubsection{Metrics}\label{subsubsection:metrics}
For a chosen query and cut-off value of $t\in (0,1)$, we  select all pairs $(i,j)$ for which $\pi_{i,j}\ge t$. This leads to a true positive rate $\TPR_t=|\{(i,j):\pi_{i,j}\ge t\} \cap S |/|S|$, where $S:=\{(i,j): A_{i,j}=1\}$ is the set of correct answers (for example the set of true direct causal effects for the query \texttt{isParent}). The corresponding false positive rate  is $\FPR_t=|\{(i,j):\pi_{i,j}\ge t\} \cap S^c |/|S^c|$, with $S^c:=\{(i,j): A_{i,j}=0\}$. The four metrics we consider are as follows.
\begin{enumerate}[(i)]
\item {\bf AOC.} The standard area-under-curve (AUC) measures the area below the graph  $(\FPR_t,\TPR_t)\in [0,1]^2$ as $t$ is varied between $0$ and $1$. Under random guessing, the area is 0.5 in expectation and the optimal values is 1. Here, to make rates comparable, we look at the area-above-curve defined as AOC $= 1-$AUC, such that low values are preferable.
\item {\bf Equal-error-rate (E-ER).} The equal-error-rate measures the false-negative rate  $\FNR_t = 1-\TPR_{t}$ at the cutoff $t$ where it equals the false-positive-rate $\FPR_t$, that is, for the value $t\in (0,1)$ for which $1-\TPR_{t}=\FPR_{t}$. The advantage over AOC is that it is a real error rate and is also identical whether we look at the missing edges or at the true edges. For random guessing, the expected value is 0.5 and does not depend on the sparsity of the graph.
\item {\bf No-false-positives-error-rate (NFP-ER).} The no-false-positives-error-rate  measures the false negative rate $\FNR_t = 1-\TPR_{t}$ for the minimal cutoff $t$ at which $\FPR_{t}=0$, that is, for the largest number of selections under the constraint that not a single false positive occurs. The expected value under random guessing depends on the sparsity of the graph.
\item {\bf No-false-negatives-error-rate (NFN-ER).} The no-false-negatives-detection-rate measures the false-positive rate $\FPR_{t} = 1 - \mathrm{TNR}_t$ for the maximally large cutoff $t$ at which $\FNR_{t}=0$, that is, for the smallest number of selections possible that not a single false negative occurs. The expected value under random guessing depends on the sparsity of the graph.
\end{enumerate}
All four metrics are designed so that lower values are better.

\subsection{Results}\label{subsec:evalresults}

Below, we mostly present results for the \texttt{isAncestor} query and the metric E-ER. Results for other queries and metrics are similar in nature.

\begin{figure}[h!]
\begin{center}
\includegraphics[width=0.6\textwidth]{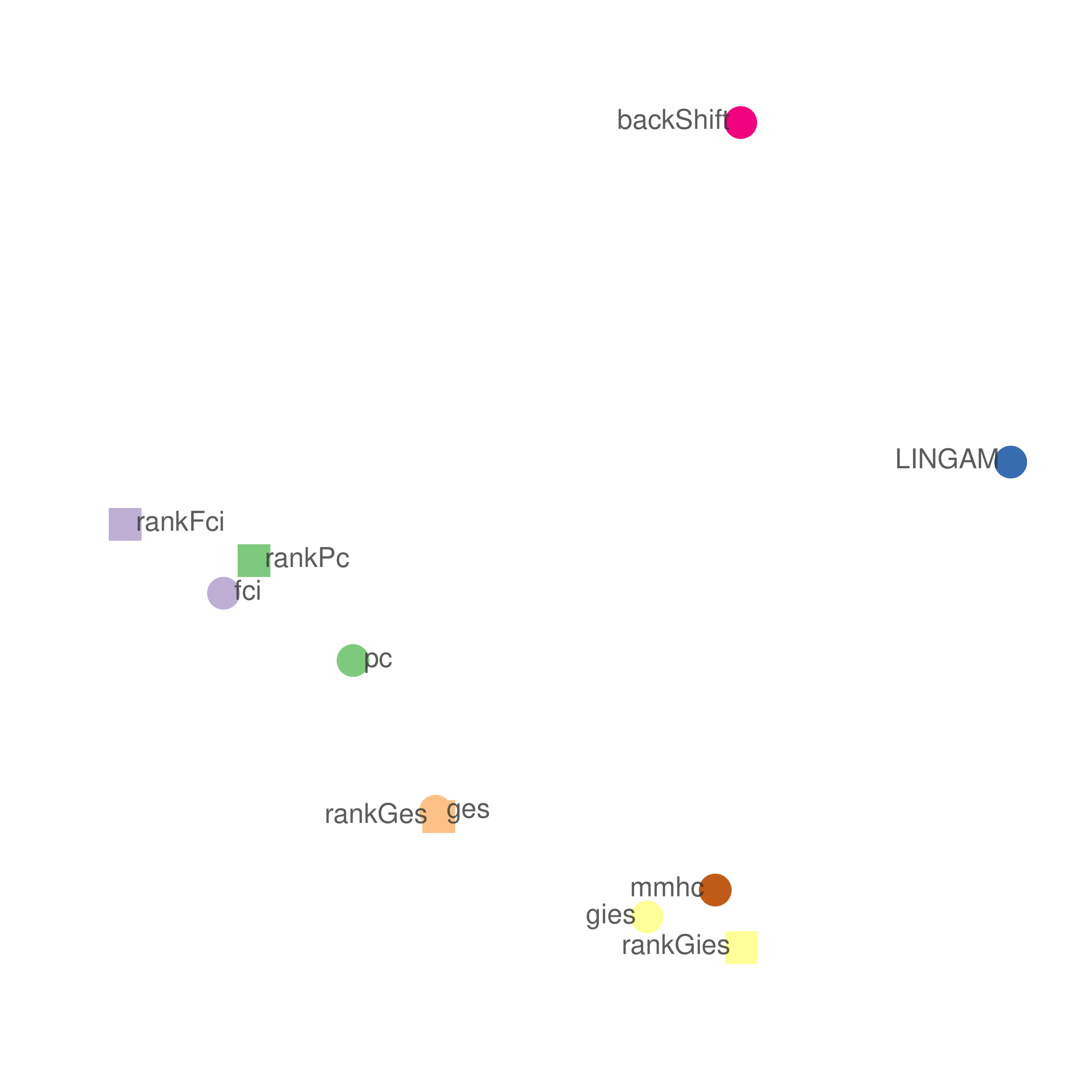}
\caption{A multi-dimensional scaling visualization of the methods
  considered.  The distance between two methods is taken to be the
  Euclidean distance between the equal-error-rate of both methods across
  all settings for the \texttt{isAncestor} query. The MDS plot uses least-squares scaling.}
\label{figMDS}
\end{center}
\end{figure}

\begin{figure}[h!]
\begin{center}
\includegraphics[width=0.9\textwidth]{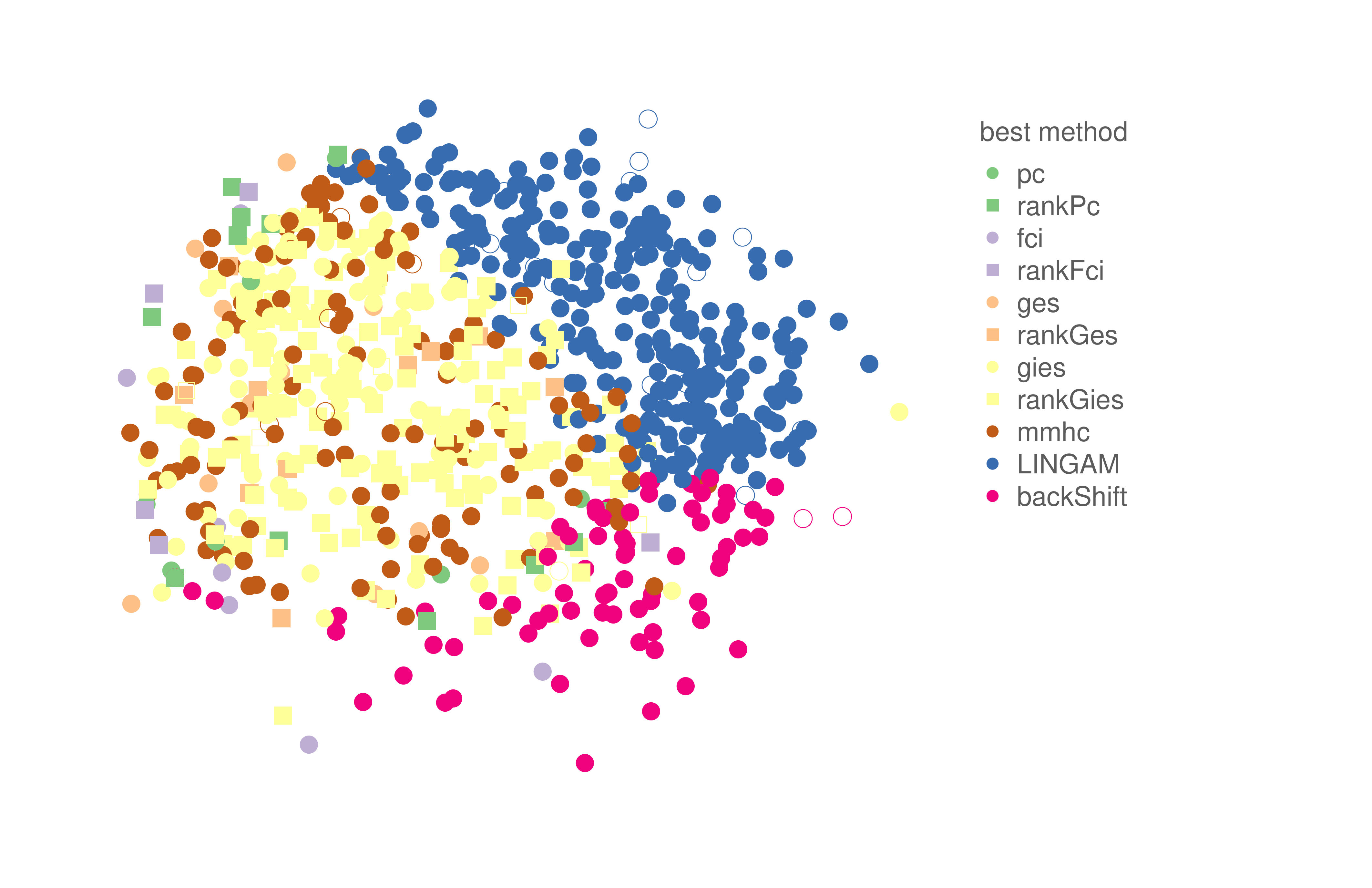}
\caption{A multi-dimensional scaling visualization of the simulation
  settings.  The distance between two simulation settings is taken to be the
  correlation distance between the equal-error-rate of both simulation settings across
  all methods for the \texttt{isAncestor} query. Each setting is shown as a sample point with color coding
for the best performing method. A filled symbol indicates that the
performance metric was
smaller than 0.3 and an  un-filled symbol that it was
above. MDS uses least-squares scaling.  }
\label{figMDSsettings}
\end{center}
\end{figure}

\subsubsection{Multi-dimensional scaling}

For each simulation setting and each method, we compute the
equal-error-rate for the \texttt{isAncestor}
query. This yields a (nr of simulation settings) $\times$ (nr of methods) matrix with E-ER values. The Euclidean distance between two columns in this matrix is a distance between methods. Similarly, the Euclidean distance between two rows in the matrix is a distance between simulation settings.

Figure~\ref{figMDS} shows an MDS plot based on distances between the methods, using least-squares scaling.
We see that the rank-based methods  rankFCI, rankPC, rankGES and
 rankGIES
are close to their counterparts FCI, PC, GES and
  GIES. It is somewhat unexpected that MMHC is closer to GIES and rankGIES than to PC and GES.
  The two methods that have the largest average distance to the other methods are LINGAM and  BACKSHIFT. This is perhaps
expected as these methods are of a very different nature than the other methods.

Figure~\ref{figMDSsettings} shows an MDS plot based on distances between the simulation settings, again using least-squares scaling. Thus, each point in the plot now corresponds to a simulation setting. The points are colored according to the best performing method.
We see that the regions where either LINGAM or BACKSHIFT are optimal are
relatively well separated, while the regions where GIES, MMHC, PC, GES,
FCI or their rank-based versions are optimal, do not show a clear
separation, as perhaps already expected from the previous result in Figure~\ref{figMDS}.

\subsubsection{Pairwise comparisons}
\begin{table}[!htbp] \centering
  \caption{A pairwise comparison. Each column shows the percentage
    of settings where methods were better by a  margin of at least 0.1
    in the equal-error-rate compared to method in the given column. For example, LINGAM beats PC in $14\%$ of the settings, while PC beats LINGAM by the given marge in $29\%$ of the settings.
    There is no
    globally dominant algorithm and a block-structure among related
    algorithms is visible.
    }
  \label{tabComp}
\begin{tabular}{@{\extracolsep{0pt}} cccccccccccc}
\\[-1.8ex]\hline
\hline \\[-1.8ex]
 &\rot{PC }&\rot{rankPC }&\rot{FCI }&\rot{rankFCI }&\rot{GES
                                                          }&\rot{
                                                             rankGES
                                                             }&\rot{
                                                                GIES
                                                                }&\rot{
                                                                   rankGIES
                                                                   }&\rot{
                                                                      MMHC
                                                                      }&\rot{
                                                                         LINGAM
                                                                         }&\rot{
                                                                            BACKSHIFT
  }\\
\hline \\[-1.8ex]
PC & $0$ & $6$ & $10$ & $16$ & $1$ & $1$ & $1$ & $0$ & $0$ & $14$ & $20$ \\
rankPC & $0$ & $0$ & $9$ & $10$ & $1$ & $2$ & $0$ & $0$ & $0$ & $11$ & $17$ \\
FCI & $1$ & $9$ & $0$ & $5$ & $1$ & $1$ & $1$ & $0$ & $0$ & $11$ & $17$ \\
rankFCI & $0$ & $1$ & $0$ & $0$ & $1$ & $1$ & $0$ & $0$ & $0$ & $10$ & $16$ \\
GES & $5$ & $15$ & $16$ & $23$ & $0$ & $0$ & $1$ & $0$ & $0$ & $16$ & $26$ \\
rankGES & $6$ & $15$ & $16$ & $24$ & $0$ & $0$ & $1$ & $0$ & $1$ & $16$ & $25$ \\
GIES & $18$ & $29$ & $26$ & $35$ & $10$ & $11$ & $0$ & $0$ & $2$ & $25$ & $35$ \\
rankGIES & $26$ & $36$ & $34$ & $44$ & $17$ & $17$ & $4$ & $0$ & $1$ & $27$ & $38$ \\
MMHC & $21$ & $33$ & $30$ & $40$ & $16$ & $17$ & $5$ & $0$ & $0$ & $23$ & $36$ \\
LINGAM & $29$ & $34$ & $34$ & $38$ & $27$ & $27$ & $19$ & $14$ & $14$ & $0$ & $31$ \\
BACKSHIFT & $18$ & $23$ & $24$ & $29$ & $16$ & $16$ & $9$ & $5$ & $7$
& $13$ & $0$ \\
\hline \\[-1.8ex]
\end{tabular}
\end{table}

Next, we investigate whether there are methods that dominate the others.
 We compare the equal-error-rate across all different settings in
Table~\ref{tabComp}.  It is apparent that no such dominance is visible among
different pairs of methods.
A block-structure is  visible, however, with similar groups as in Figure \ref{figMDS}.
One block is formed by the
constraint-based methods \{PC,
rankPC, FCI, rankFCI\}: the equal-error-rate of constraint-based
methods  is hardly ever substantially different. The second block is
formed by the score-based approaches
\{GES, rankGIES\} and the third given by the extensions and hybrid
methods \{GIES, rankGIES, MMHC\}. This latter block is of interest as MMHC makes fewer
assumptions about the available data and does not need to know where
interventions occurred.
LINGAM and BACKSHIFT, on the other hand, do not fit nicely into any
block in the empirical comparison  and  are
more  orthogonal to the other algorithms in that they perform
substantially better \emph{and} substantially worse in many settings, if
compared to the other approaches.

\begin{table}[!htbp]
 \begin{center}
 \caption{Marginal rank correlations  between
   equal-error-rate performance (for the \texttt{isAncestor} query) on the one hand
   and parameters settings on the other hand (shown only if  absolute
   value exceeds 0.1, multiplied by 100 and rounded to the nearest
   multiple of 5).   A  positive value for $p$ indicates, for example,
    that the method becomes less successful with increasing $p$.
}
  \label{table:cor}
\begin{small}
\begin{tabular}{@{\extracolsep{-1pt}} cccccccccccc}
\\[-1.8ex]\hline
\hline \\[-1.8ex]
 &\rot{ PC }&\rot{ rankPC }&\rot{ FCI }&\rot{ rankFCI }&\rot{ GES }&\rot{ rankGES }&\rot{ GIES }&\rot{ rankGIES }&\rot{ MMHC }& \rot{LINGAM} & \rot{BACKSHIFT} \\
\hline \\[-1.8ex]
n &  & $15$ &  & $10$ &  &  &  &  &  &  & $$-$15$ \\
p & $45$ & $45$ & $25$ & $25$ & $40$ & $35$ & $35$ & $40$ & $45$ & $40$ & $75$ \\
$\mathrm{df}_\varepsilon$ &  &  &  &  &  &  &  &  &  & $15$ &  \\
$\rho_\varepsilon$ & $50$ & $60$ & $55$ & $60$ & $55$ & $55$ & $65$ & $50$ & $50$ & $35$ &  \\
$\omega$ & $10$ & $10$ & $10$ & $10$ & $15$ & $10$ & $20$ & $15$ & $10$ &  & $20$ \\
$p_s$ & $20$ & $15$ & $20$ & $15$ & $25$ & $25$ & $25$ & $30$ & $30$ & $15$ & $25$ \\
do-interv &  &  & $$-$10$ & $$-$10$ &  &  &  &  &  &  &  \\
$n_I$ &  &  &  &  &  &  &  &  &  &  &  \\
$\sigma_Z$ & $$-$35$ & $$-$25$ & $$-$35$ & $$-$30$ & $$-$35$ & $$-$35$ & $$-$25$ & $$-$35$ & $$-$30$ &  & $$-$30$ \\
cyclic &  &  & $$-$15$ & $$-$15$ &  &  &  &  &  & $35$ &  \\
$w_c$ &  &  & $$-$15$ & $$-$15$ &  &  &  &  &  & $35$ &  \\
nonlinear &  &  &  &  &  &  &  &  &  & $20$ &  \\
\hline \\[-1.8ex]
\end{tabular}

\end{small}
\end{center}
\end{table}

\begin{figure}[h]
\begin{center}
\includegraphics[width=0.75\textwidth]{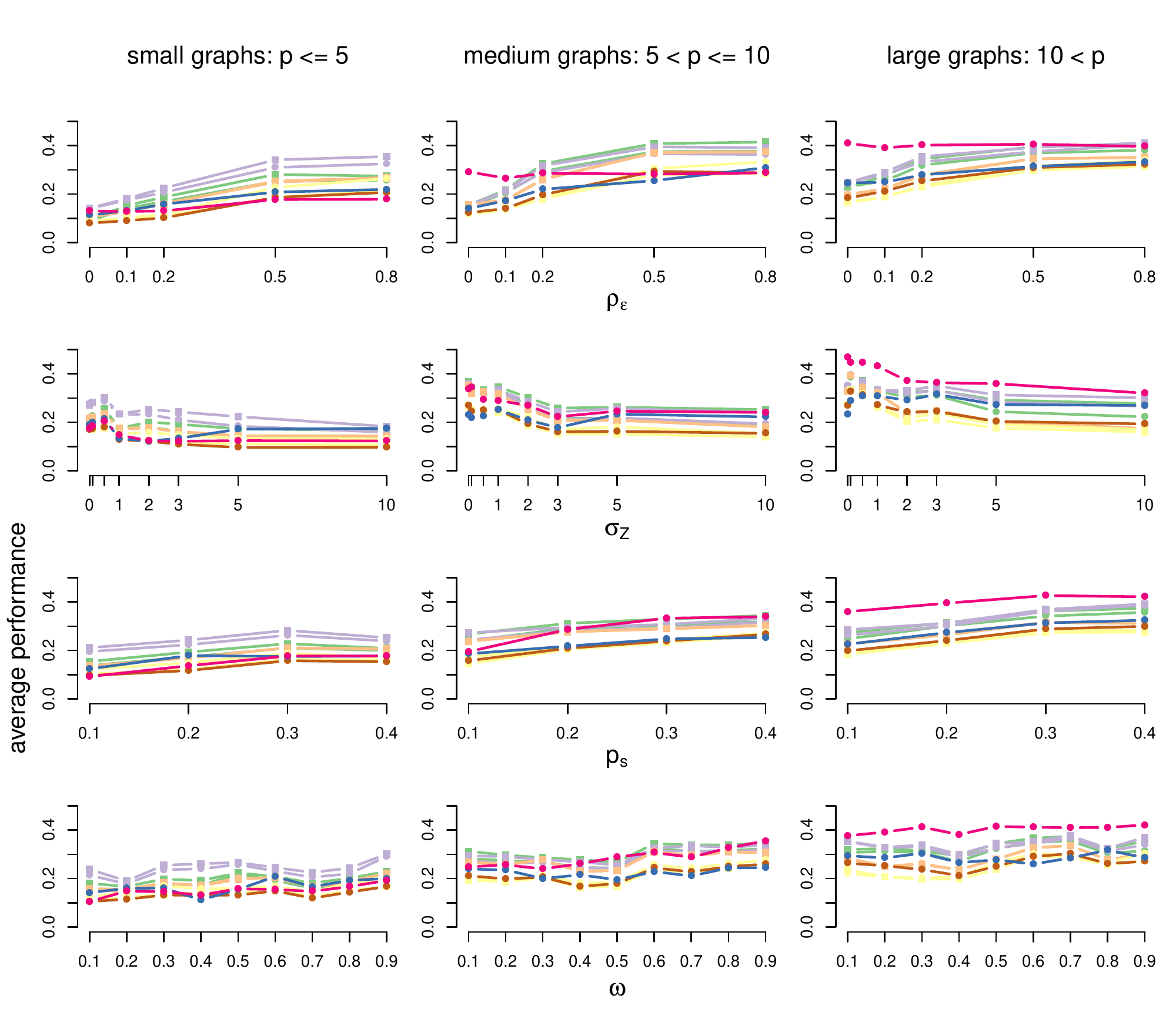}
\caption{The average equal-error-rate for the \texttt{isAncestor} query, for each method as a function for the
  four most important parameters (besides the number of
  variables $p$). The left column shows results for small graphs ($p\le 5$),
  the middle column intermediate graphs ($5< p \le 10$), and the right column for
  large graphs ($p>10$). The color coding is identical to previous
  plots. }
\label{figInter}
\end{center}
\end{figure}

\begin{figure}[h]
\begin{center}
\includegraphics[width=0.75\textwidth]{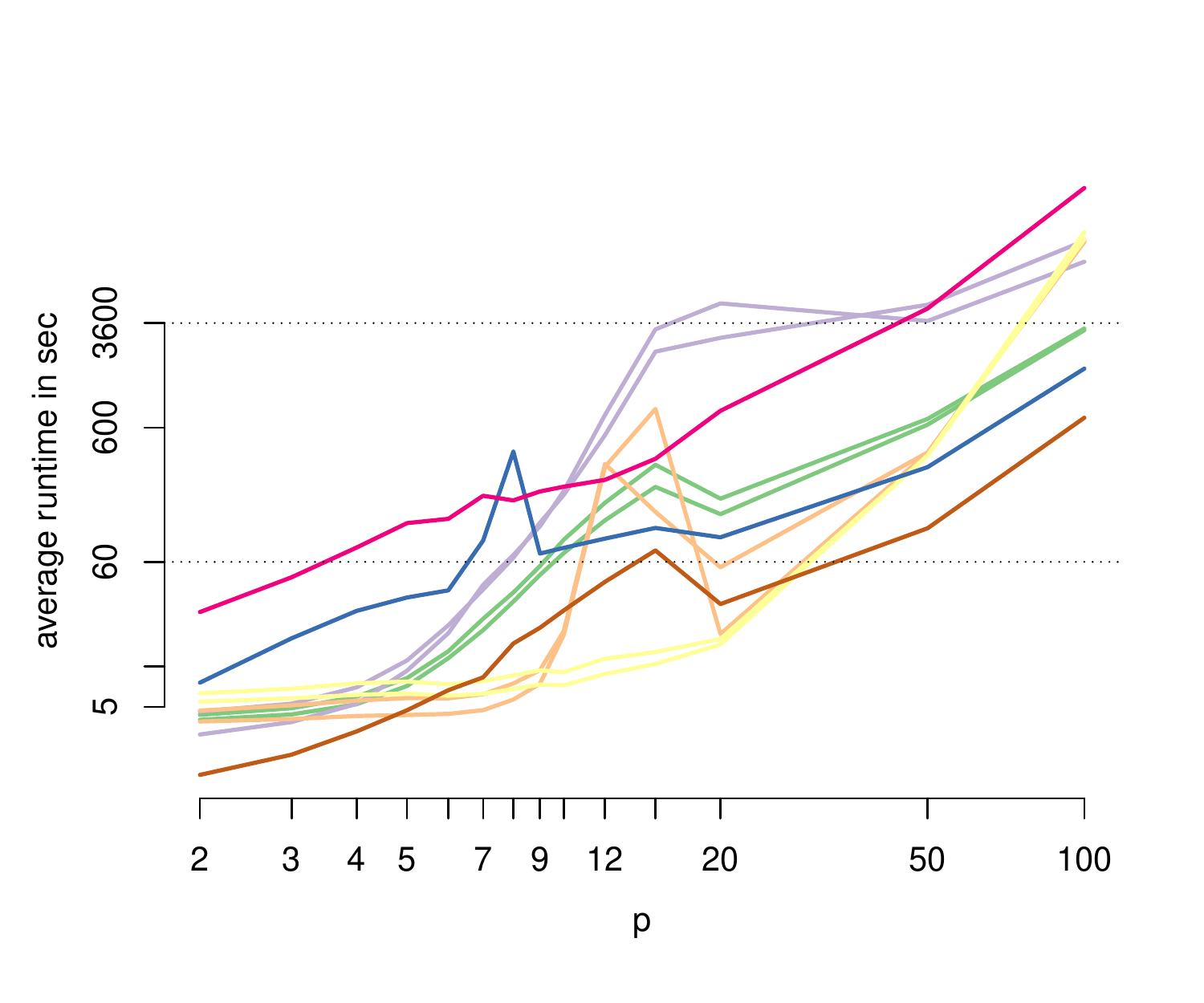}
\caption{The average runtime in seconds of each method on a
  logarithmic scale as a function of the number of variables $p$ on a
  logarithmic scale. A minute and one hour is shown as horizontal bars. The time includes the stability ranking. A single run is  faster by a factor of 100 for all methods. (A single run of BACKSHIFT already includes ten subsamples.)}
\label{figInter}
\end{center}
\end{figure}

\subsubsection{Which causal graphs can be estimated well?}
Which graphs can be estimated by some or all methods? To start
answering the question, we show in Table~\ref{table:cor} the rank correlation between the
equal-error-rate for the \texttt{isAncestor} query and parameter settings for all methods. We see that
the number of variables $p$ and the strength of the
hidden variables $\rho_\varepsilon$ show the highest correlations. In both cases the correlation is positive, indicating that increased $p$ or $\rho_\varepsilon$ leads to higher equal-error-rates. Other parameters that show substantial correlations are $\omega$, $p_s$ and $\sigma_Z$. For $\omega$ and $p_s$ we again see positive correlations, indicating that large noise contributions and denser graphs are associated with higher equal-error-rates. The correlation with $\sigma_Z$ is negative for all methods except for LINGAM. While it is expected that BACKSHIFT benefits from strong interventions, the benefit for for example PC and FCI is unexpected.

We note that the strong effect of $\rho_\varepsilon$ can be explained by the fact that we created a correlation $\rho_\varepsilon$  between all pairs of noise variables. It is not surprising that this has a larger impact than adding for example a single cycle to the graph (which only seems to substantially affect the performance of LINGAM).

Figure~\ref{figInter} shows the average
equal-error-rate for the \texttt{isAncestor} query for each method as
a function of the simulation parameters $\rho_\varepsilon$, $\omega$, $p_s$ and $\sigma_Z$ as identified from Table \ref{table:cor}, split according to the
number of variables $p$ in the graph (small, medium-sized and large
graphs). Again, we see that the size of the graph $p$ and the strength of the hidden variables $\rho_\varepsilon$ have the strongest effect on performance, with the exception that BACKSHIFT is not much affected by $\rho_\varepsilon$ (but which is also perhaps less competitive in the absence of latent confounding). The strength of the
interventions, the sparsity of the graph and the signal-to-noise ratio
also affect the average performance but perhaps to a lesser
extent.

Some other observations:
\begin{enumerate}[(a)]
\item The most surprising outcome is perhaps that
the number of samples $n$ has only a very weak influence on the
success despite it being varied between a few hundred and twenty
thousand.
\item Sparser graphs with fewer edges are consistently easier to
  estimate with all methods than dense graphs.
\item Less  heavy tails in the error distribution have an adverse
  effect on the performance of LINGAM only, as it makes use of higher
  moments.
 LINGAM is also most affected when each variable undergoes a nonlinear transformation. 
\item A cycle in the graph again has a detrimental effect on LINGAM
  (which is likely different in the version of LINGAM that allows for
  cycles \citep{lacerda2012discovering}).
\end{enumerate}

\begin{figure}[h!]
\begin{center}
\begin{minipage}{0.4\textwidth}
        \includegraphics[width=0.99\textwidth]{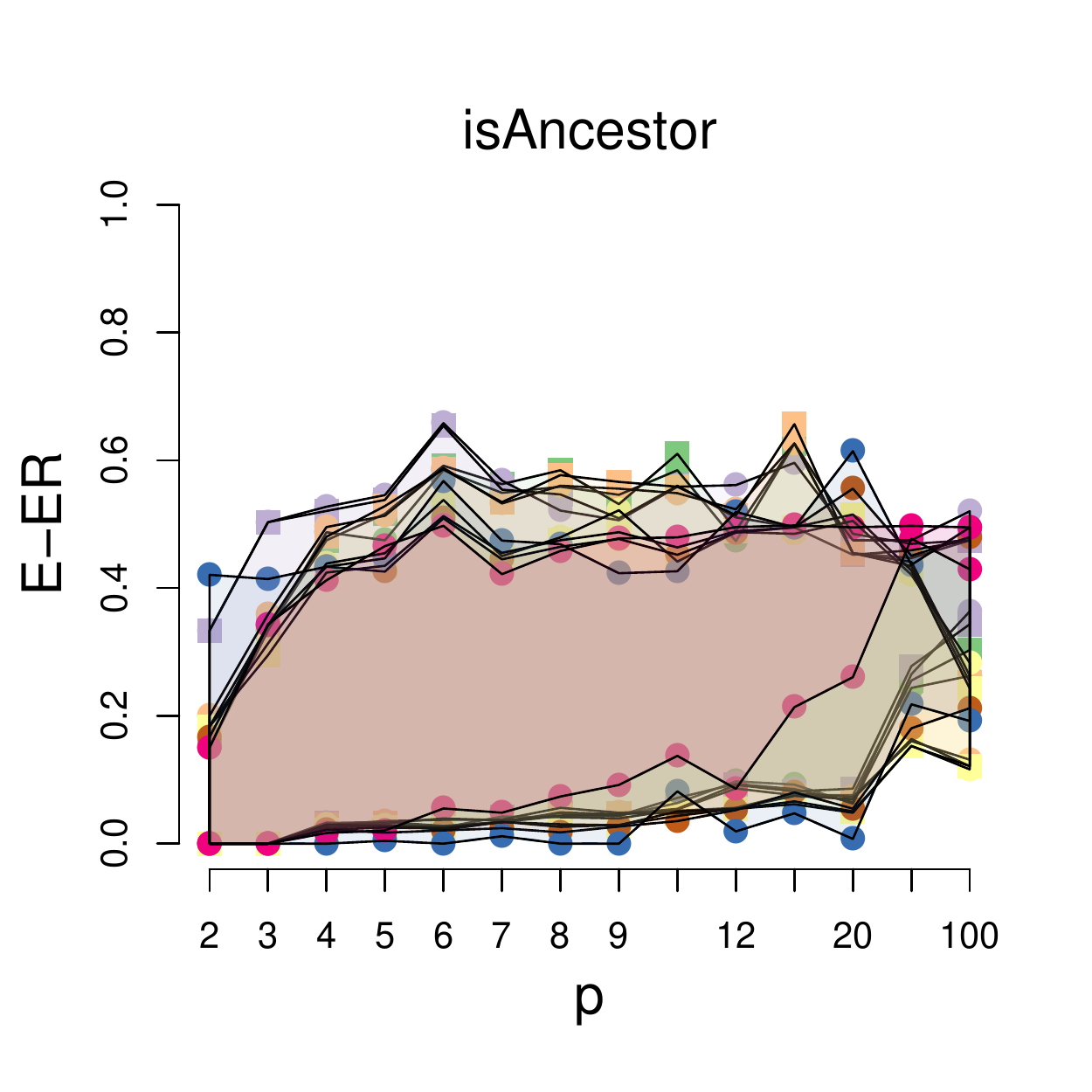}
\end{minipage}
\begin{minipage}{0.4\textwidth}
        \includegraphics[width=0.99\textwidth]{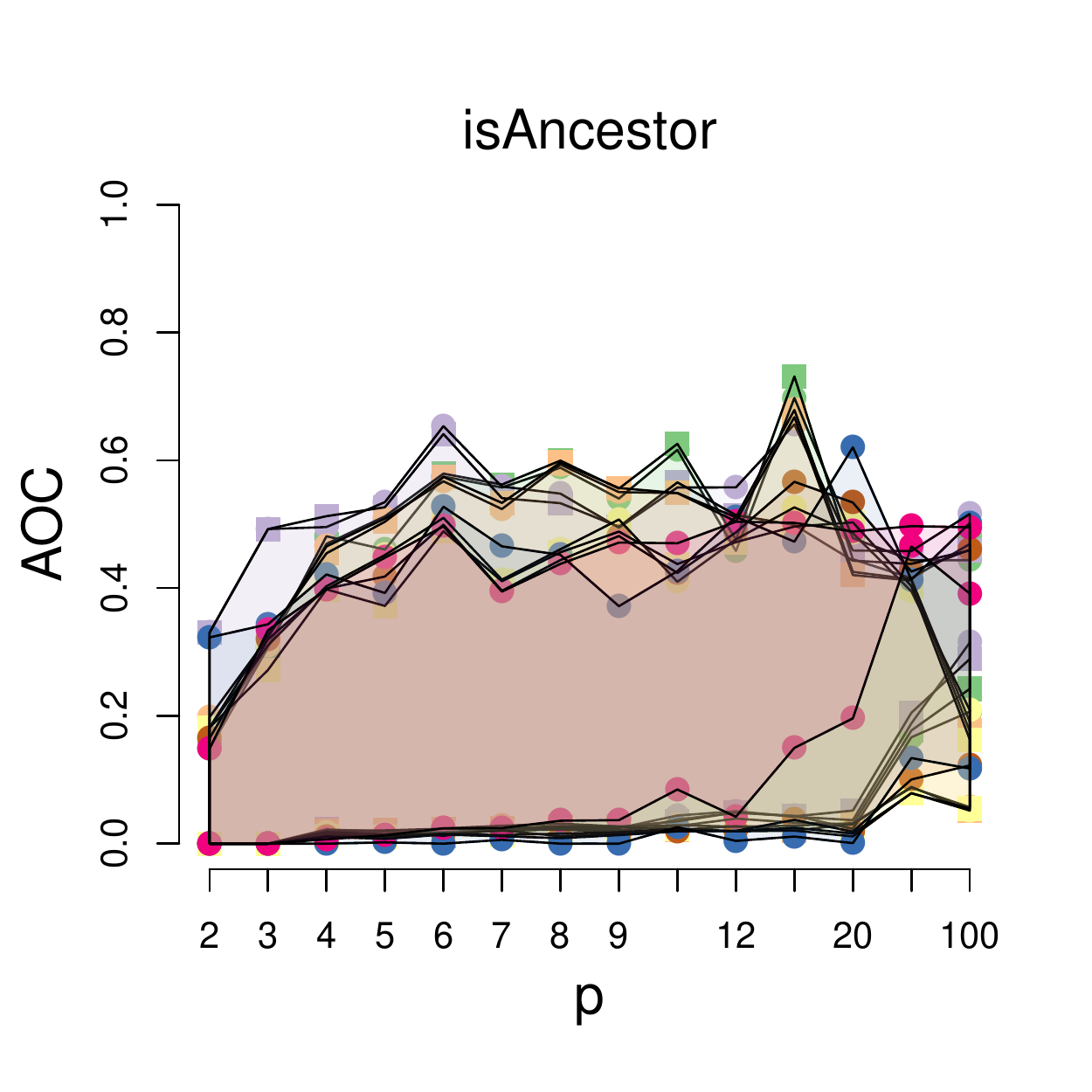}
\end{minipage}

\begin{minipage}{0.4\textwidth}
        \includegraphics[width=0.99\textwidth]{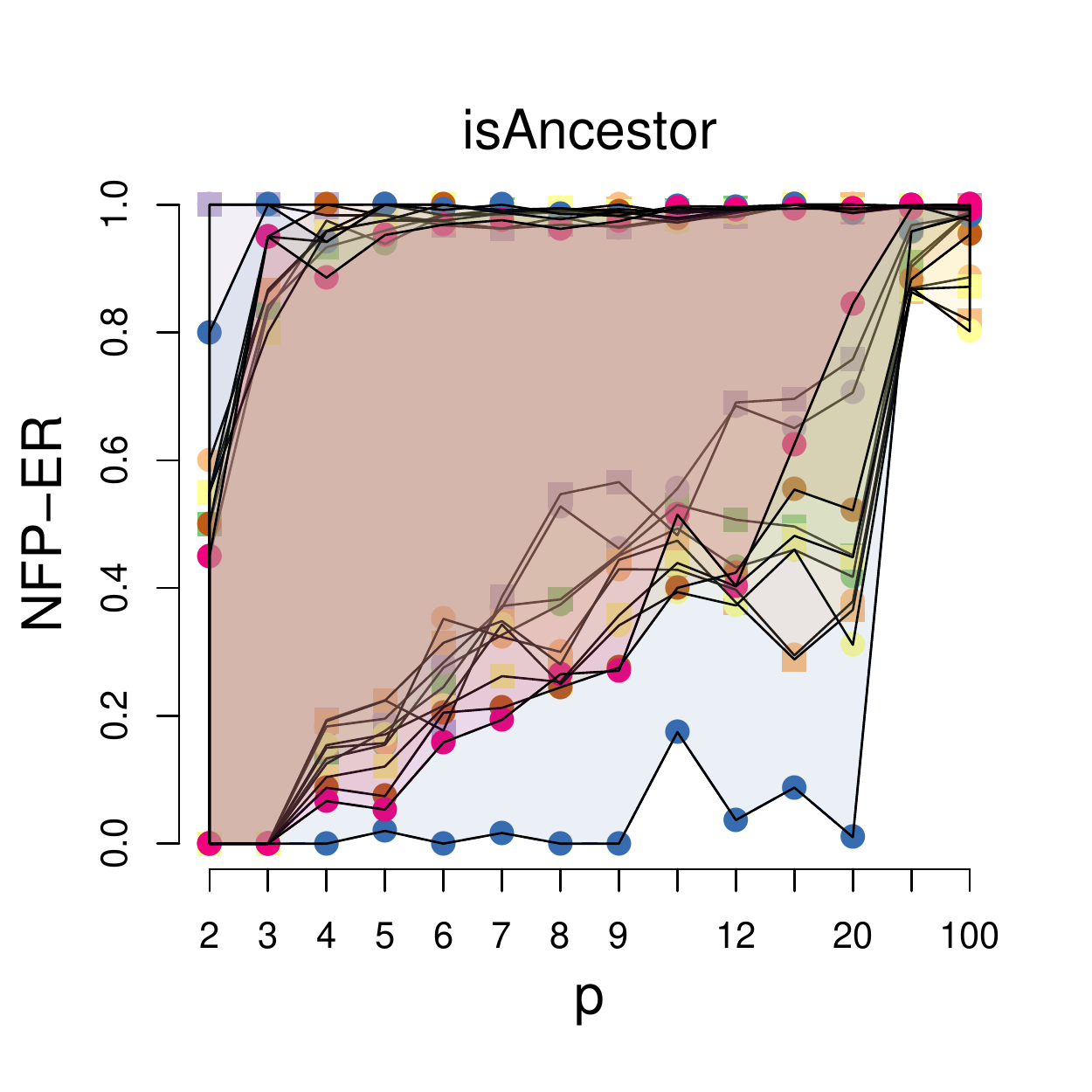}
\end{minipage}
\begin{minipage}{0.4\textwidth}
        \includegraphics[width=0.99\textwidth]{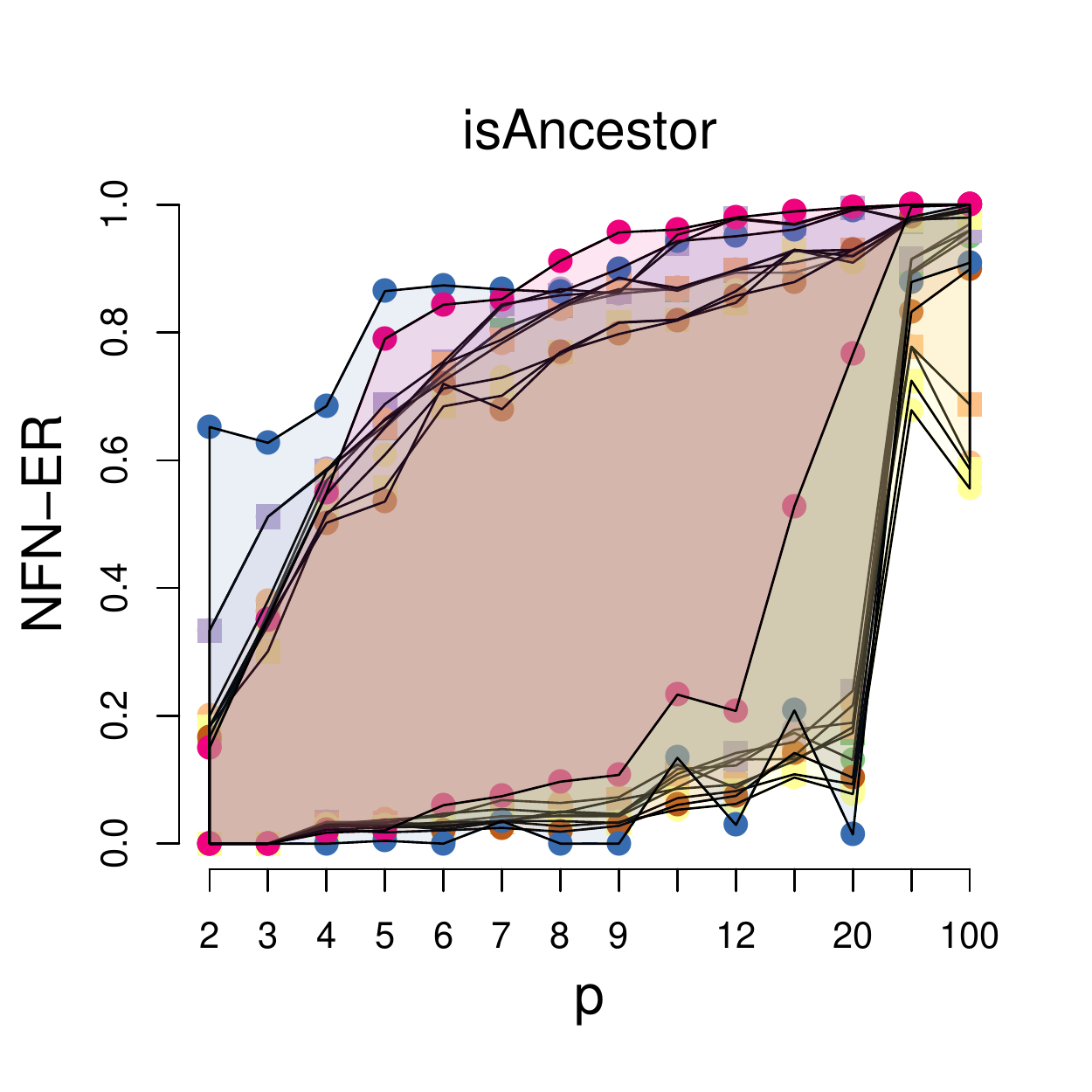}
\end{minipage}

\begin{minipage}{0.4\textwidth}
        \includegraphics[width=0.99\textwidth]{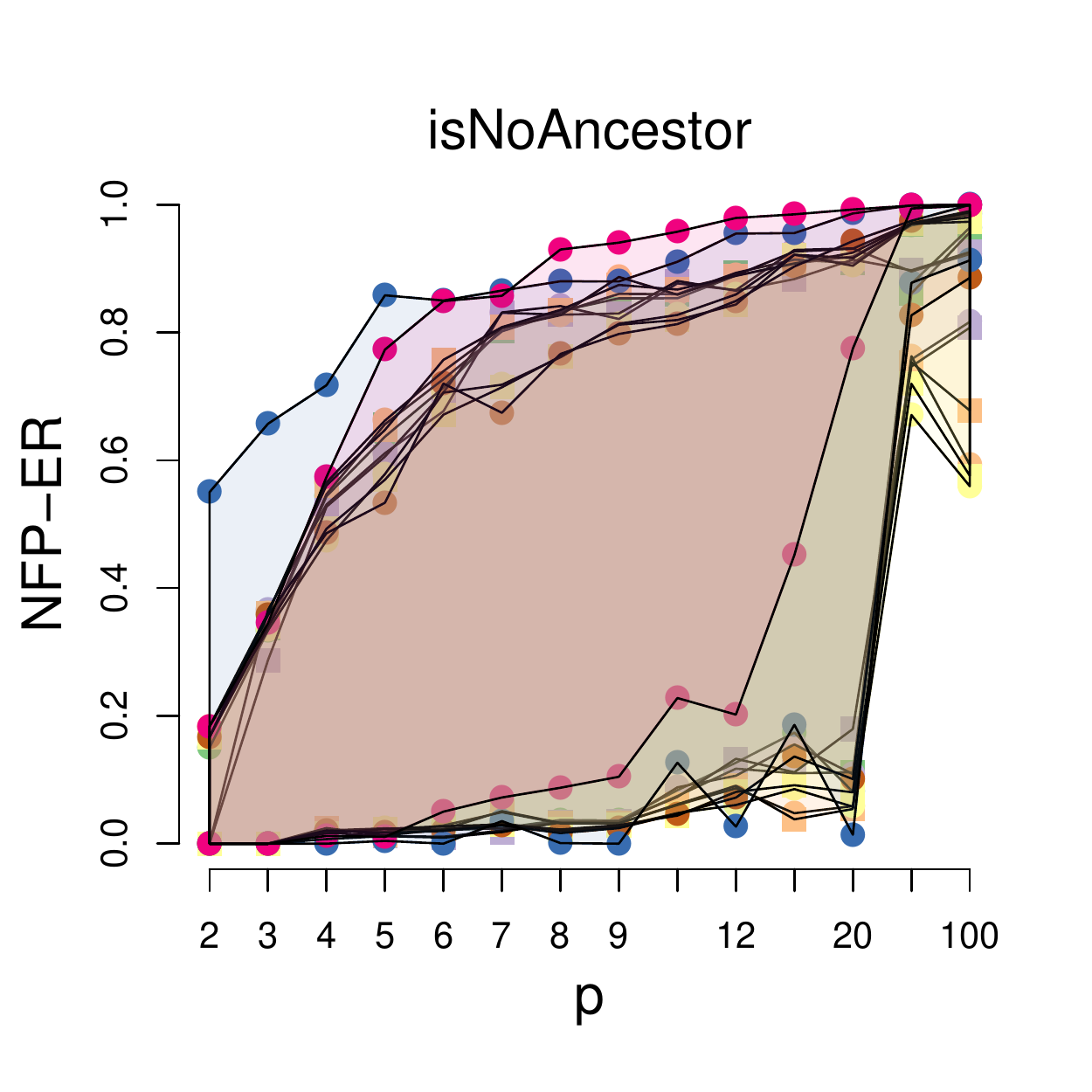}
\end{minipage}
\begin{minipage}{0.4\textwidth}
        \includegraphics[width=0.99\textwidth]{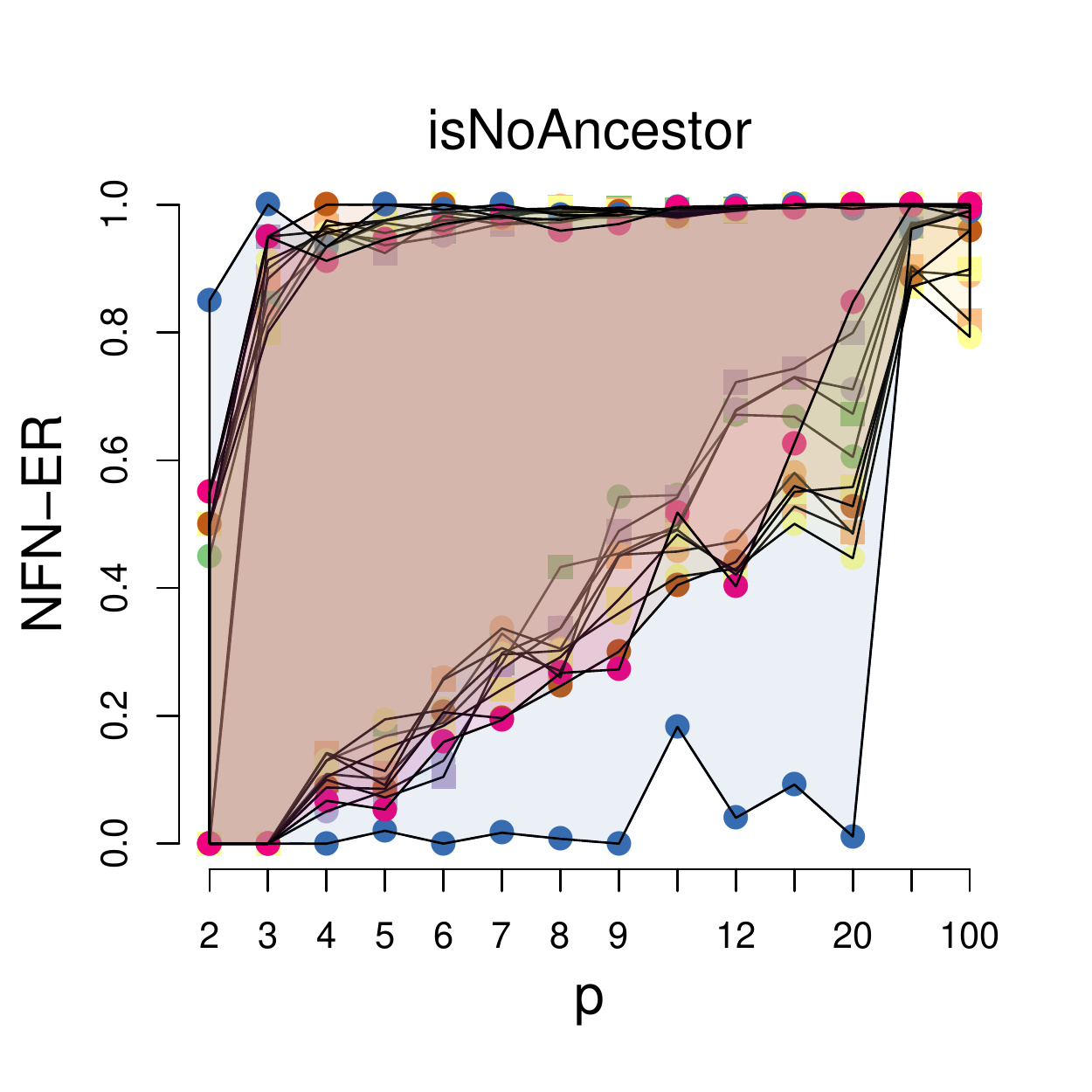}
\end{minipage}

\caption{The range of equal-error rate (E-ER) for all methods as a
  function of the number of variables $p$ for the \texttt{isAncestor}
  query (top left). Top right shows the same  for the area-above-curve
  (AOC), while second row shows the  no-false-positives-error-rate (NFP-ER) and
  no-false-negatives-error-rate (NFN-ER).
The last row contains the corresponding plots to the second row but
for the \texttt{isNoAncestor} query.
}
\label{figInterNo}
\end{center}
\end{figure}

\subsubsection{Bounds on performance}
The outcome of the simulations show a large degree of variation. To further investigate the
role of the number of variables $p$,
we show in
Figure~\ref{figInterNo} the bounds of the  performance
as a function of $p$ for the \texttt{isAncestor} query. Specifically, for each value
of $p$, we consider the range of the four considered metrics when varying all other
parameters for each method and show the lower and upper bounds in the
figure.

The upper bounds show the worst performance across all parameters while
holding $p$ constant. It can be compared to the expected value under
random guessing which is 0.5 for the E-ER and AOC metrics and 1 for
NFP-ER and NFN-ER.

The lower bound reveals in contrast the error rates in the best
setting for a given $p$.
The metric NFP-ER seems more difficult to keep at reasonable
levels than NFN-ER, with the exception of LINGAM which has very small
values of NFP-ER in some settings up to $p\approx 20$. The NFN-ER rate
is typically lower than NFP-ER as there are typically more
non-ancestral pairs in the graphs (due to not connected components for
example) as ancestral pairs. This is confirmed by the third row of
panels in Figure~\ref{figInterNo} which shows the error rates for the
\texttt{isNoAncestor} query. Here the roles of NFN-ER and NFP-ER
are reversed due to the relative abundance of non-ancestral pairs.

\FloatBarrier

\section{DISCUSSION}
\label{sec:discussion}

We have tried to give a contemporaneous overview of structure learning
for causal models that are available in \textsf{R} and conducted an extensive empirical comparison. It
is noteworthy that we found a clustering of methods into constraint-based,
score-based, and other approaches that do not fall neatly into these
categories. Methods
from the same class behave empirically very similar. We also tried to
quantify to what extent methods are negatively or positively affected
by various parameters such as the size of the graph to learn, sparsity
and strength of hidden variables. The most important parameters in our set-up are the
size of the graph $p$ and the strength of the hidden variables $\rho_\varepsilon$.
An easily accessible interface to
all methods is contributed as \textsf{R}-package
\texttt{CompareCausalNetworks}.

The results suggest that more efficient algorithms would be desirable, both from a computational
and from a statistical point-of-view.
As it stands, the success of the algorithms depends on both the assumptions made
about the data generating process (and how accurate these assumptions
are) and the specific implementation details of each
algorithm. It would be worthwhile if the relative importance of these
two factors could be separated better by more modular estimation
methods and perhaps more work on worst-case bounds. These latter
bounds would allow to quantify to what extent the empirically poor statistical
scalability is inherent to the problem or  a consequence of choices
made in  the considered
algorithms.

\newpage
\section{APPENDIX}
\subsection{Considered tuning parameter configurations}\label{app:subsubsec:methods_pars}
All methods were run through the interface offered by the \texttt{CompareCausalNetworks} package \citep{CCN-r-pkg}. Below we also indicate the \textsf{R} packages from which the \texttt{CompareCausalNetworks} package calls the respective methods.

\paragraph*{backShift}
Code available from the \textsf{R} package \texttt{backShift} \citep{backshift-pkg}.
\begin{itemize}
	\item[--] \texttt{covariance} $\in \{ \texttt{TRUE, FALSE} \}$ 
	\item[--] \texttt{ev} $\in \{  0.1,0.25,0.5\}\cdot \dims$ 
	\item[--] \texttt{threshold} $ = 0.75$
	\item[--] \texttt{nsim} $ = 10$
	\item[--] \texttt{sampleSettings} $=1/sqrt(2)$
	\item[--] \texttt{sampleObservations} $=1/sqrt(2)$
	\item[--] \texttt{nodewise = TRUE}
	\item[--] \texttt{tolerance} $=10^{-4}$
\end{itemize}

\paragraph*{GES and rankGES}
Code available from the \textsf{R} packages \texttt{pcalg} \citep{KalischEtAl12} (GES) and  \texttt{CompareCausalNetworks} (rankGES).

\begin{itemize}
	\item[--] \texttt{phase = 'turning'}
	\item[--] \texttt{score = GaussL0penObsScore}
	\item[--] $\lambda \in \{0.05\log{n}, 0.5\log{n}, 5\log{n} \}$ 
	\item[--] \texttt{adaptive = "none"}
	\item[--] \texttt{maxDegree = integer(0)}
\end{itemize}

\paragraph*{GIES and rankGIES}
Code available from the \textsf{R} packages \texttt{pcalg} \citep{KalischEtAl12} (GIES) and  \texttt{CompareCausalNetworks} (rankGIES).
\begin{itemize}
	\item[--] \texttt{phase = 'turning'}
	\item[--] \texttt{score = GaussL0penObsScore}
	\item[--] $\lambda \in \{0.05\log{n}, 0.5\log{n}, 5\log{n} \}$ 
	\item[--] \texttt{adaptive = "none"}
	\item[--] \texttt{maxDegree = integer(0)}
\end{itemize}

\paragraph*{FCI and rankFCI}
Code available from the \textsf{R} packages \texttt{pcalg} \citep{KalischEtAl12} (FCI) and  \texttt{CompareCausalNetworks} (rankFCI).
\begin{itemize}
	\item[--] \texttt{conservative = FALSE} and \texttt{maj.rule = FALSE} 
	\item[--] \texttt{conservative = TRUE} and \texttt{maj.rule = FALSE} 
	\item[--] \texttt{conservative = FALSE} and \texttt{maj.rule = TRUE}
	\item[--] \texttt{alpha} $\in \{0.001, 0.01, 0.1\}$ 
	\item[--] \texttt{indepTest = gaussCItest}
	\item[--] \texttt{skel.method = "stable"}
	\item[--] \texttt{m.max = Inf}
	\item[--] \texttt{pdsep.max = Inf}
	\item[--] \texttt{rules = rep(TRUE,10)}
	\item[--] \texttt{NAdelete = TRUE}
	\item[--] \texttt{doPdsep = TRUE}
	\item[--] \texttt{biCC = FALSE}
\end{itemize}


\paragraph*{MMHC}
Code available from the \textsf{R} package \texttt{bnlearn} \citep{bnlearn-pkg}.

\begin{itemize}
	\item[--] $\lambda \in \{0.05\log{n}, 0.5\log{n}, 5\log{n} \}$ 	
	\item[--] \texttt{alpha} $\in \{0.001, 0.01, 0.1\}$ 
	\item[--] \texttt{whitelist = NULL}
	\item[--] \texttt{blacklist = NULL}
	\item[--] \texttt{test = NULL} -- corresponds to correlation
	\item[--] \texttt{score = NULL} -- corresponds to BIC
	\item[--] \texttt{B = NULL}
	\item[--] \texttt{restart = 0}
	\item[--] \texttt{perturb = 1}
	\item[--] \texttt{max.iter = Inf}
	\item[--] \texttt{optimized = TRUE}
	\item[--] \texttt{strict = FALSE}
\end{itemize}

\paragraph*{PC and Rank PC}
Code available from the \textsf{R} packages \texttt{pcalg} \citep{KalischEtAl12} (PC) and  \texttt{CompareCausalNetworks} (rankPC).
\begin{itemize}
	\item[--] \texttt{conservative = FALSE} and \texttt{maj.rule = FALSE} 
	\item[--] \texttt{conservative = TRUE} and \texttt{maj.rule = FALSE} 
	\item[--] \texttt{conservative = FALSE} and \texttt{maj.rule = TRUE}
	\item[--] \texttt{alpha} $\in \{0.001, 0.01, 0.1\}$ 
	\item[--] \texttt{indepTest = gaussCItest}
	\item[--] \texttt{NAdelete = TRUE}
	\item[--] \texttt{m.max = Inf}
	\item[--] \texttt{u2pd = "relaxed"}
	\item[--] \texttt{skel.method = "stable"}
	\item[--] \texttt{solve.confl = FALSE} 
\end{itemize}

\subsection{Simulation settings}\label{app:subsubsec:sim_settings}
The results in this work are based on 842 unique simulation settings. The tables below show for each parameter in the data generation scheme how many settings were generated for each considered value for the given parameter.

\paragraph*{Sample size}\mbox{}\\

\begin{tabular}{@{\extracolsep{1pt}} ccccc} 
$\samp$ & 500 & 2000 & 5000 & 10000 \\ \hline \\[-1.8ex] 
\# of settings & 231 & 200 & 217 & 194 \\ 
\end{tabular} 

\paragraph*{Number of variables}\mbox{}\\

\begin{tabular}{@{\extracolsep{1pt}} ccccccccccccccc} 
$\dims$ & 2 & 3 & 4 & 5 & 6 & 7 & 8 & 9 & 10 & 12 & 15 & 20 & 50 & 100 \\  \hline \\[-1.8ex] 
\# of settings & 71 & 89 & 84 & 77 & 62 & 60 & 74 & 68 & 62 & 76 & 60 & 43 &  8 &  8 \\ 
\end{tabular} 

\paragraph*{Edge density parameter}\mbox{}\\

\begin{tabular}{@{\extracolsep{1pt}} ccccc} 
$\sparse$ & 0.1 & 0.2 & 0.3 & 0.4 \\ \hline \\[-1.8ex]
\# of settings & 202 & 226 & 200 & 214 \\ 
\end{tabular} 

\paragraph*{Number of settings}\mbox{}\\

\begin{tabular}{@{\extracolsep{1pt}} cccc} 
$\numberInt$ & 3 & 4 & 5 \\ \hline \\[-1.8ex]
\# of settings & 271 & 275 & 296 \\ 
\end{tabular} 

\paragraph*{Intervention type}\mbox{}\\

\begin{tabular}{@{\extracolsep{1pt}} ccc} 
& shift intervention & do-intervention \\ \hline \\[-1.8ex]
\# of settings & 417 & 425 \\ 
\end{tabular} 

\paragraph*{Strength of the interventions}\mbox{}\\

\begin{tabular}{@{\extracolsep{1pt}} ccccccccc} 
$\strengthInt$ & 0 & 0.1 & 0.5 & 1 & 2 & 3 & 5 & 10 \\ \hline \\[-1.8ex]
\# of settings & 111 & 105 & 102 & 105 & 106 &  98 & 116 &  99 \\ 
\end{tabular} 

\paragraph*{Degrees of freedom of the noise distribution}\mbox{}\\

\begin{tabular}{@{\extracolsep{1pt}} ccccccc} 
$\dfNoise$ & 2 & 3 & 5 & 10 & 20 & 100 \\ \hline \\[-1.8ex]
\# of settings & 140 & 136 & 147 & 140 & 144 & 135 \\ 
\end{tabular} 

\paragraph*{Strength of hidden variables}\mbox{}\\

\begin{tabular}{@{\extracolsep{1pt}} cccccc} 
$\rhoNoise$ & 0 & 0.1 & 0.2 & 0.5 & 0.8 \\ \hline \\[-1.8ex]
\# of settings & 161 & 164 & 166 & 179 & 172 \\ 
\end{tabular} 

\paragraph*{Proportion of variance from noise }\mbox{}\\

\begin{tabular}{@{\extracolsep{1pt}} cccccccccc} 
$\snrPar$ & 0.1 & 0.2 & 0.3 & 0.4 & 0.5 & 0.6 & 0.7 & 0.8 & 0.9 \\ \hline \\[-1.8ex]
\# of settings & 116 &  85 &  88 & 110 &  85 &  91 &  82 &  91 &  94 \\ 
\end{tabular} 

\paragraph*{Settings with cycles}\mbox{}\\
\begin{tabular}{@{\extracolsep{1pt}} ccc} 
& no cycles & cycles \\ \hline \\[-1.8ex]
\# of settings & 576 & 266 \\ 
\end{tabular} 

\paragraph*{Strength of cycle}\mbox{}\\

\begin{tabular}{@{\extracolsep{1pt}} ccccccc} 
$\strengthCycle$  & 0 & 0.1 & 0.25 & 0.5 & 0.75 & 0.9 \\ \hline \\[-1.8ex]
\# of settings & 576 &  56 &  51 &  50 &  55 &  54 \\ 
\end{tabular} 

\paragraph*{Settings with model misspecification}\mbox{}\\

\begin{tabular}{@{\extracolsep{1pt}} ccc} 
& no model misspecification & model misspecification \\ \hline \\[-1.8ex]
\# of settings & 715 & 127 \\ 
\end{tabular} 


\begin{thebibliography}{53}
\providecommand{\natexlab}[1]{#1}
\providecommand{\url}[1]{\texttt{#1}}
\expandafter\ifx\csname urlstyle\endcsname\relax
  \providecommand{\doi}[1]{doi: #1}\else
  \providecommand{\doi}{doi: \begingroup \urlstyle{rm}\Url}\fi

\bibitem[Ali et~al.(2009)Ali, Richardson, and Spirtes]{AliRichardsonSpirtes09}
R.~Ayesha Ali, Thomas~S. Richardson, and P.~Spirtes.
\newblock Markov equivalence for ancestral graphs.
\newblock \emph{Ann. Stat.}, 37:\penalty0 2808--2837, 2009.

\bibitem[Andersson et~al.(1997)Andersson, Madigan, and Perlman]{Andersson1997}
S.A. Andersson, D.~Madigan, and M.D. Perlman.
\newblock A characterization of {M}arkov equivalence classes for acyclic
  digraphs.
\newblock \emph{Annals of Statistics}, 25:\penalty0 505--541, 1997.

\bibitem[Angrist et~al.(1996)Angrist, Imbens, and
  Rubin]{angrist1996identification}
J.~D. Angrist, G.~W. Imbens, and D.~B. Rubin.
\newblock Identification of causal effects using instrumental variables.
\newblock \emph{Journal of the American Statistical Association}, 91:\penalty0
  444--455, 1996.

\bibitem[Chickering(2002{\natexlab{a}})]{Chickering02-CPDAG}
D.~M. Chickering.
\newblock Learning equivalence classes of {B}ayesian-network structures.
\newblock \emph{Journal of Machine Learning Research}, 2:\penalty0 445--498,
  2002{\natexlab{a}}.

\bibitem[Chickering(2002{\natexlab{b}})]{Chickering2002}
D.~M. Chickering.
\newblock Optimal structure identification with greedy search.
\newblock \emph{Journal of Machine Learning Research}, 3:\penalty0 507--554,
  2002{\natexlab{b}}.

\bibitem[Cho et~al.(2014)Cho, Kim, Kim, Kweon, Kim, Bae, and Kim]{ChoEtAl14}
S.~W. Cho, S.~Kim, Y.~Kim, J.~Kweon, H.~S. Kim, S.~Bae, and J.-S. Kim.
\newblock Analysis of off-target effects of {CRISPR}/{C}as-derived {RNA}-guided
  endonucleases and nickases.
\newblock \emph{Genome Research}, 24:\penalty0 132--141, 2014.

\bibitem[Claassen et~al.(2013)Claassen, Mooij, and Heskes]{Claassen2013}
T.~Claassen, J.~M. Mooij, and T.~Heskes.
\newblock Learning sparse causal models is not {NP}-hard.
\newblock In \emph{Proceedings of the 29th Annual Conference on {U}ncertainty
  in {A}rtificial {I}ntelligence ({UAI})}, 2013.

\bibitem[Colombo and Maathuis(2014)]{ColomboMaathuis14}
D.~Colombo and M.~H. Maathuis.
\newblock Order-independent constraint-based causal structure learning.
\newblock \emph{J. Mach. Learn. Res.}, 15:\penalty0 3741--3782, 2014.

\bibitem[Colombo et~al.(2012)Colombo, Maathuis, Kalisch, and
  Richardson]{Colombo2012}
D.~Colombo, M.~H. Maathuis, M.~Kalisch, and T.~S. Richardson.
\newblock Learning high-dimensional directed acyclic graphs with latent and
  selection variables.
\newblock \emph{Annals of Statistics}, 40:\penalty0 294--321, 2012.

\bibitem[Comon(1994)]{Comon94}
P.~Comon.
\newblock Independent component analysis, a new concept?
\newblock \emph{Signal processing}, 36:\penalty0 287--314, 1994.

\bibitem[Cooper and Yoo(1999)]{cooper1999causal}
G.~Cooper and C.~Yoo.
\newblock Causal discovery from a mixture of experimental and observational
  data.
\newblock In \emph{Proceedings of the 15th Annual Conference on {U}ncertainty
  in {A}rtificial {I}ntelligence ({UAI})}, pages 116--125, 1999.

\bibitem[Dawid(2000)]{dawid2000causal}
A.~P. Dawid.
\newblock Causal inference without counterfactuals.
\newblock \emph{Journal of the American Statistical Association}, 95:\penalty0
  407--424, 2000.

\bibitem[Didelez(2017)]{didelez2017causalconcepts}
V.~Didelez.
\newblock \emph{Handbook of Graphical Models}, chapter Causal Concepts and
  Graphical Models.
\newblock Chapman \& Hall/CRC, 2017.
\newblock To appear.

\bibitem[Drton and Maathuis(2017)]{DrtonMaathuis17}
M.~Drton and M.~H. Maathuis.
\newblock Structure learning in graphical modeling.
\newblock \emph{Annual Review of Statistics and Its Application}, 4:\penalty0
  365--393, 2017.

\bibitem[Eaton and Murphy(2007)]{Eaton2007}
D.~Eaton and K.~P. Murphy.
\newblock Exact {B}ayesian structure learning from uncertain interventions.
\newblock In \emph{Proceedings of the 11th International Conference on
  Artificial Intelligence and Statistics ({AISTATS})}, pages 107--114, 2007.

\bibitem[Frisch(1938)]{Frisch38}
R.~Frisch.
\newblock Autonomy of economic relations: Statistical versus theoretical
  relations in economic macrodynamics.
\newblock Paper given at League of Nations. Reprinted in D.F. Hendry and M.S.
  Morgan (1995), The Foundations of Econometric Analysis, Cambridge University
  Press, 1938.

\bibitem[Haavelmo(1944)]{Haavelmo1944}
T.~Haavelmo.
\newblock The probability approach in econometrics.
\newblock \emph{Econometrica}, 12:\penalty0 S1--S115 (supplement), 1944.

\bibitem[Harris and Drton(2013)]{harrisdrton13}
N.~Harris and M.~Drton.
\newblock P{C} algorithm for nonparanormal graphical models.
\newblock \emph{Journal of Machine Learning Research}, 14:\penalty0 3365--3383,
  2013.

\bibitem[Hauser and B{\"u}hlmann(2012)]{Hauser2012}
A.~Hauser and P.~B{\"u}hlmann.
\newblock Characterization and greedy learning of interventional {M}arkov
  equivalence classes of directed acyclic graphs.
\newblock \emph{Journal of Machine Learning Research}, 13:\penalty0 2409--2464,
  2012.

\bibitem[Heinze-Deml(2017)]{backshift-pkg}
C.~Heinze-Deml.
\newblock \emph{backShift: Learning Causal Cyclic Graphs from Unknown Shift
  Interventions}, 2017.
\newblock URL \url{https://github.com/christinaheinze/backShift}.
\newblock R package version 0.1.4.1.

\bibitem[Heinze-Deml and Meinshausen(2017)]{CCN-r-pkg}
C.~Heinze-Deml and N.~Meinshausen.
\newblock \emph{{C}ompare{C}ausal{N}etworks: Interface to Diverse Estimation
  Methods of Causal Networks}, 2017.
\newblock URL \url{https://github.com/christinaheinze/CompareCausalNetworks}.
\newblock {R} package version 0.1.6.

\bibitem[Hoyer et~al.(2008)Hoyer, Shimizu, Kerminen, and
  Palviainen]{Hoyer2008b}
P.~O. Hoyer, S.~Shimizu, A.~J. Kerminen, and M.~Palviainen.
\newblock Estimation of causal effects using linear non-{G}aussian causal
  models with hidden variables.
\newblock \emph{Int. J. Approx. Reasoning}, 49:\penalty0 362--378, 2008.

\bibitem[Hyttinen et~al.(2012)Hyttinen, Eberhardt, and Hoyer]{Hyttinen2012}
A.~Hyttinen, F.~Eberhardt, and P.~O. Hoyer.
\newblock Learning linear cyclic causal models with latent variables.
\newblock \emph{Journal of Machine Learning Research}, 13:\penalty0 3387--3439,
  2012.

\bibitem[Imbens(2014)]{imbens2014instrumental}
G.~Imbens.
\newblock Instrumental variables: An econometrician’s perspective.
\newblock \emph{Statistical Science}, 29:\penalty0 323--358, 2014.

\bibitem[Kalisch and B\"{u}hlmann(2007)]{Kalisch2007}
M.~Kalisch and P.~B\"{u}hlmann.
\newblock Estimating high-dimensional directed acyclic graphs with the
  {PC}-algorithm.
\newblock \emph{Journal of Machine Learning Research}, 8:\penalty0 613--636,
  2007.

\bibitem[Kalisch et~al.(2012)Kalisch, M\"achler, Colombo, Maathuis, and
  B\"uhlmann]{KalischEtAl12}
M.~Kalisch, M.~M\"achler, D.~Colombo, M.~H. Maathuis, and P.~B\"uhlmann.
\newblock Causal inference using graphical models with the {R} package
  \texttt{pcalg}.
\newblock \emph{Journal of Statistical Software}, 47\penalty0 (11):\penalty0
  1--26, 2012.

\bibitem[Lacerda et~al.(2008)Lacerda, Spirtes, Ramsey, and
  Hoyer]{lacerda2012discovering}
G.~Lacerda, P.~Spirtes, J.~Ramsey, and P.O. Hoyer.
\newblock Discovering cyclic causal models by independent components analysis.
\newblock In \emph{Proceedings of the 24th Conference on Uncertainty in
  Artificial Intelligence (UAI)}, pages 366--374, 2008.

\bibitem[Lauritzen(1996)]{Lauritzen1996}
S.~L. Lauritzen.
\newblock \emph{Graphical Models}.
\newblock Oxford University Press, New York, USA, 1996.

\bibitem[Maathuis et~al.(2009)Maathuis, Kalisch, and B\"uhlmann]{Maathuis2009}
M.~H. Maathuis, M.~Kalisch, and P.~B\"uhlmann.
\newblock Estimating high-dimensional intervention effects from observational
  data.
\newblock \emph{Annals of Statistics}, 37:\penalty0 3133--3164, 2009.

\bibitem[Maathuis et~al.(2010)Maathuis, Colombo, Kalisch, and
  B\"uhlmann]{Maathuis2010}
M.~H. Maathuis, D.~Colombo, M.~Kalisch, and P.~B\"uhlmann.
\newblock Predicting causal effects in large-scale systems from observational
  data.
\newblock \emph{Nature {M}ethods}, 7:\penalty0 247--248, 2010.

\bibitem[Nandy et~al.(2017{\natexlab{a}})Nandy, Hauser, and
  Maathuis]{NandyEtAl17}
P.~Nandy, A.~Hauser, and M.~H. Maathuis.
\newblock High-dimensional consistency in score-based and hybrid structure
  learning.
\newblock 2017{\natexlab{a}}.
\newblock arXiv:1507.02608.

\bibitem[Nandy et~al.(2017{\natexlab{b}})Nandy, Maathuis, and
  Richardson]{Nandy2014}
P.~Nandy, M.~H. Maathuis, and T.~S. Richardson.
\newblock Estimating the effect of joint interventions from observational data
  in high-dimensional settings.
\newblock \emph{Annals of Statistics}, 45:\penalty0 647--674,
  2017{\natexlab{b}}.

\bibitem[Pearl(2009)]{Pearl2009}
J.~Pearl.
\newblock \emph{Causality: Models, Reasoning, and Inference}.
\newblock Cambridge University Press, New York, USA, 2nd edition, 2009.

\bibitem[Peters et~al.(2016)Peters, B{\"u}hlmann, and
  Meinshausen]{peters2015causal}
J.~Peters, P.~B{\"u}hlmann, and N.~Meinshausen.
\newblock Causal inference using invariant prediction: identification and
  confidence intervals.
\newblock \emph{Journal of the Royal Statistical Society, Series B},
  78:\penalty0 947--1012, 2016.

\bibitem[{R Core Team}(2017)]{R}
{R Core Team}.
\newblock \emph{R: A Language and Environment for Statistical Computing}.
\newblock R Foundation for Statistical Computing, Vienna, Austria, 2017.
\newblock URL \url{https://www.R-project.org/}.

\bibitem[Richardson and Robins(2013)]{richardson2013single}
T.~Richardson and J.~M. Robins.
\newblock Single world intervention graphs ({SWIGs}): A unification of the
  counterfactual and graphical approaches to causality.
\newblock \emph{Center for the Statistics and the Social Sciences, University
  of Washington Series. Working Paper 128, 30 April 2013}, 2013.

\bibitem[Richardson and Spirtes(1999)]{Richardson1999}
T.~Richardson and P.~Spirtes.
\newblock Automated discovery of linear feedback models.
\newblock In C.~Glymour and G.F. Cooper, editors, \emph{Computation, Causation,
  and Discovery}, pages 253--304. MIT Press, 1999.

\bibitem[Richardson and Spirtes(2002)]{Richardson2002}
T.~Richardson and P.~Spirtes.
\newblock Ancestral graph {M}arkov models.
\newblock \emph{Annals of Statistics}, 30:\penalty0 962--1030, 2002.

\bibitem[Robins(1986)]{Robins1986}
J.~M. Robins.
\newblock A new approach to causal inference in mortality studies with a
  sustained exposure period -- application to control of the healthy worker
  survivor effect.
\newblock \emph{Mathematical Modelling}, 7:\penalty0 1393 -- 1512, 1986.

\bibitem[Rothenh{\"a}usler et~al.(2015)Rothenh{\"a}usler, Heinze, Peters, and
  Meinshausen]{rothenhausler2015backshift}
D.~Rothenh{\"a}usler, C.~Heinze, J.~Peters, and N.~Meinshausen.
\newblock {backShift}: Learning causal cyclic graphs from unknown shift
  interventions.
\newblock In \emph{{A}dvances in {N}eural {I}nformation {P}rocessing {S}ystems
  28 ({NIPS})}, pages 1513--1521, 2015.

\bibitem[Rubin(2005)]{rubin2005causal}
D.~B. Rubin.
\newblock Causal inference using potential outcomes.
\newblock \emph{Journal of the American Statistical Association}, 100:\penalty0
  322--331, 2005.

\bibitem[Scutari(2010)]{bnlearn-pkg}
M.~Scutari.
\newblock Learning bayesian networks with the {bnlearn} {R} package.
\newblock \emph{Journal of Statistical Software}, 35\penalty0 (3):\penalty0
  1--22, 2010.
\newblock URL \url{http://www.jstatsoft.org/v35/i03/}.

\bibitem[Shimizu et~al.(2006)Shimizu, Hoyer, Hyv\"{a}rinen, and
  Kerminen]{Shimizu2006}
S.~Shimizu, P.~O. Hoyer, A.~Hyv\"{a}rinen, and A.J. Kerminen.
\newblock A linear non-{G}aussian acyclic model for causal discovery.
\newblock \emph{Journal of Machine Learning Research}, 7:\penalty0 2003--2030,
  2006.

\bibitem[Shimizu et~al.(2011)Shimizu, Inazumi, Sogawa, Hyv\"{a}rinen, Kawahara,
  Washio, Hoyer, and Bollen]{Shimizu2011}
S.~Shimizu, T.~Inazumi, Y.~Sogawa, A.~Hyv\"{a}rinen, Y.~Kawahara, T.~Washio,
  P.~O. Hoyer, and K.~Bollen.
\newblock Direct{LiNGAM}: A direct method for learning a linear non-{G}aussian
  structural equation model.
\newblock \emph{Journal of Machine Learning Research}, 12:\penalty0 1225--1248,
  2011.

\bibitem[Spirtes et~al.(1999)Spirtes, Meek, and
  Richardson]{SpirtesMeekRichardson99}
P.~Spirtes, C.~Meek, and T.S. Richardson.
\newblock \emph{Computation, Causation and Discovery}, chapter An algorithm for
  causal inference in the presence of latent variables and selection bias,
  pages 211--252.
\newblock MIT Press, 1999.

\bibitem[Spirtes et~al.(2000)Spirtes, Glymour, and Scheines]{Spirtes2000}
P.~Spirtes, C.~Glymour, and R.~Scheines.
\newblock \emph{Causation, Prediction, and Search}.
\newblock MIT Press, Cambridge, USA, 2nd edition, 2000.

\bibitem[Stekhoven et~al.(2012)Stekhoven, Moraes, Sveinbj\"ornsson, Hennig,
  Maathuis, and B\"uhlmann]{Stekhoven2012}
D.J. Stekhoven, I.~Moraes, G.~Sveinbj\"ornsson, L.~Hennig, M.H. Maathuis, and
  P.~B\"uhlmann.
\newblock Causal stability ranking.
\newblock \emph{submitted}, 2012.

\bibitem[Tian and Pearl(2001)]{Tian2001}
J.~Tian and J.~Pearl.
\newblock Causal discovery from changes.
\newblock In \emph{Proceedings of the 17th Conference Annual Conference on
  Uncertainty in Artificial Intelligence ({UAI})}, pages 512--522, 2001.

\bibitem[Tsamardinos et~al.(2006)Tsamardinos, Brown, and
  Aliferis]{Tsamardinos2006}
I.~Tsamardinos, L.~E. Brown, and C.~F. Aliferis.
\newblock The max-min hill-climbing {B}ayesian network structure learning
  algorithm.
\newblock \emph{Machine Learning}, 65:\penalty0 31--78, 2006.

\bibitem[Wright(1934)]{wright34}
D.~Wright.
\newblock The method of path coefficients.
\newblock \emph{Annals of Mathematical Statistics}, 5:\penalty0 161--215, 1934.

\bibitem[Wright(1921)]{Wright1921}
S.~Wright.
\newblock Correlation and causation.
\newblock \emph{Journal of Agricultural Research}, 20:\penalty0 557--585, 1921.

\bibitem[Zhang(2008{\natexlab{a}})]{Zhang08-causal-reasoning-ancestral-graphs}
J.~Zhang.
\newblock Causal reasoning with ancestral graphs.
\newblock \emph{Journal of Machine Learning Research}, 9:\penalty0 1437--1474,
  2008{\natexlab{a}}.

\bibitem[Zhang(2008{\natexlab{b}})]{Zhang08-orientation-rules}
J.~Zhang.
\newblock On the completeness of orientation rules for causal discovery in the
  presence of latent confounders and selection bias.
\newblock \emph{Artificial Intelligence}, 172:\penalty0 1873--1896,
  2008{\natexlab{b}}.

\end{thebibliography}
\end{document}